\documentclass[aps,pra,groupedaddress,reprint,nofootinbib,longbibliography]{revtex4-1}

\usepackage{graphicx,dcolumn,bm,amssymb,tipa,textcomp,epstopdf}
\usepackage{amsmath}
\newcommand{\bra}[1]{\left\langle #1\right|}
\newcommand{\ket}[1]{\left| #1\right\rangle}
\newcommand{\braket}[2]{\left\langle
#1\vphantom{#2}\right|\left.#2\vphantom{#1}\right\rangle}
\newcommand{\ketbra}[2]{\left| #1\right\rangle\!\left\langle#2\right|}
\newcommand{\avg}[1]{\left\langle #1\right\rangle}
\newcommand{\be}[0]{\begin{equation}}
\newcommand{\ee}[0]{\end{equation}}
\newcommand{\intinf}[0]{\int_{-\infty}^{+\infty}}

\newcommand{\lra}\leftrightarrow
\newcommand{\eeqref}[1]{Eq.~(\ref{#1})}
\newcommand{\braketop}[3]{\left\langle
#1\vphantom{#2#3}\right|\left.#2\vphantom{#1#3}\right|\left.#3\vphantom{#1#2}\right\rangle}

\newcommand{\pr}[0]{{\rm pr}}

\newcommand{\de}[0] {{\rm d}}

\newcommand{\nna}[0] {\nonumber \\}
\newcommand{\nnb
}[0] {\\ \nonumber }
\newcommand{\iea}[0]{{\it et al.~}}
\newcommand{\ieac}[0]{{\it et al., }}
\newcommand{\sq}[0]{\ket{{\rm sq}_R}}
\newcommand{\tmsv}[0]{\ket{{\rm TMSV}_R}}

\begin{document}

\title{Squeezed light}

\author{A.~I.~Lvovsky}\affiliation{Institute for Quantum Information Science, University of Calgary, Calgary, Canada, T2N 1N4}\affiliation{Russian Quantum Center, 100 Novaya St., Skolkovo,
Moscow region, 143025, Russia}
\email{lvov@ucalgary.ca}

\begin{abstract}
The squeezed state of the electromagnetic field can be generated in many nonlinear optical processes and finds a wide range of applications in quantum information processing and quantum metrology. This article reviews the basic properties of single-and dual-mode squeezed light states, methods of their preparation and detection, as well as their quantum technology applications.
\end{abstract}
\date{\today}


\maketitle
\section{What is squeezed light?}
\subsection{Single-mode squeezed light}
In squeezed states of light, the noise of the electric field at certain phases falls below that of the vacuum state. This means that, when we turn on the squeezed light, we see \emph{less noise than no light at all}. This apparently paradoxical feature is a direct consequence of quantum nature of light and cannot be explained within the classical framework.

The basic idea of squeezing can be understood by considering the quantum harmonic oscillator, familiar from undergraduate quantum mechanics. Its vacuum state wavefunction in the dimensionless position basis is given by\footnote{We use convention $[\hat X,\hat P]=i$ for the quadrature observables.}
\begin{equation}\label{wfvacX} \psi_0(X)=\frac{1}{\pi^{1/4}}e^{-X^2/2},
\end{equation}
which in the momentum basis corresponds to
\begin{equation}\label{wfvacP}
\tilde\psi_0(P)=\frac 1{\sqrt{2\pi}}\intinf e^{-iPX}\psi_0(X)\de X=\frac{1}{\pi^{1/4}}e^{-P^2/2}
\end{equation}
(so the vacuum state wavefunction is the same in the position and momentum bases). The variance of the position and momentum observables in the vacuum state equals $\bra 0\Delta X^2\ket 0=\bra 0\Delta P^2\ket 0=1/2$.

The wavefunction of the \emph{squeezed-vacuum} state $\sq$ with the \emph{squeezing factor} $R>0$ is obtained from that of the vacuum state by means of scaling transformation:
\begin{equation}\label{wfsqX} \psi_R(X)=\frac{\sqrt R}{\pi^{1/4}}e^{-(R X)^2/2},
\end{equation}
and
\begin{equation}\label{wfsqP}
\tilde\psi_R(P)=\frac{1}{\pi^{1/4}\sqrt R}e^{-(P/R)^2/2}
\end{equation}
in the position and momentum bases, respectively. In this state, the variances of the  two canonical observables are
\begin{equation}\label{UncSimple}
\avg{\Delta X^2}=1/(2R^2)\quad {\rm and} \quad \avg{\Delta P^2}=R^2/2.
\end{equation}
If $R>1$, the position variance is below that of the vacuum state, so $\sq$ is \emph{position-squeezed}; for $R<1$ the state is \emph{momentum-squeezed}. In other words, if we prepare multiple copies of $\sq$, and perform a measurement of the squeezed observable on each copy, our measurement results will exhibit less variance than if we performed the same set of measurements on multiple copies of the vacuum state.

More generally, we say that a state of a single harmonic oscillator exhibits \emph{(quadrature) squeezing} if the variance of the position, momentum, or any other quadrature\footnote{The field \emph{quadrature} is observable $\hat X_\theta=\hat X\cos\theta+\hat P\sin\theta$, where $\theta$ is a real number known as \emph{quadrature angle}.} in which the state exhibits variance below $1/2$. In accordance with the uncertainty principle, both position and momentum observables, or any two quadratures associated with orthogonal angles, cannot be squeezed at the same time. For example, in state \eqref{wfsqX} the product $\avg{\Delta X^2}\avg{\Delta P^2}=1/4$ is the same as that for the vacuum state \eqref{wfvacX}. 

\begin{figure}[h]
\includegraphics[width=\columnwidth]{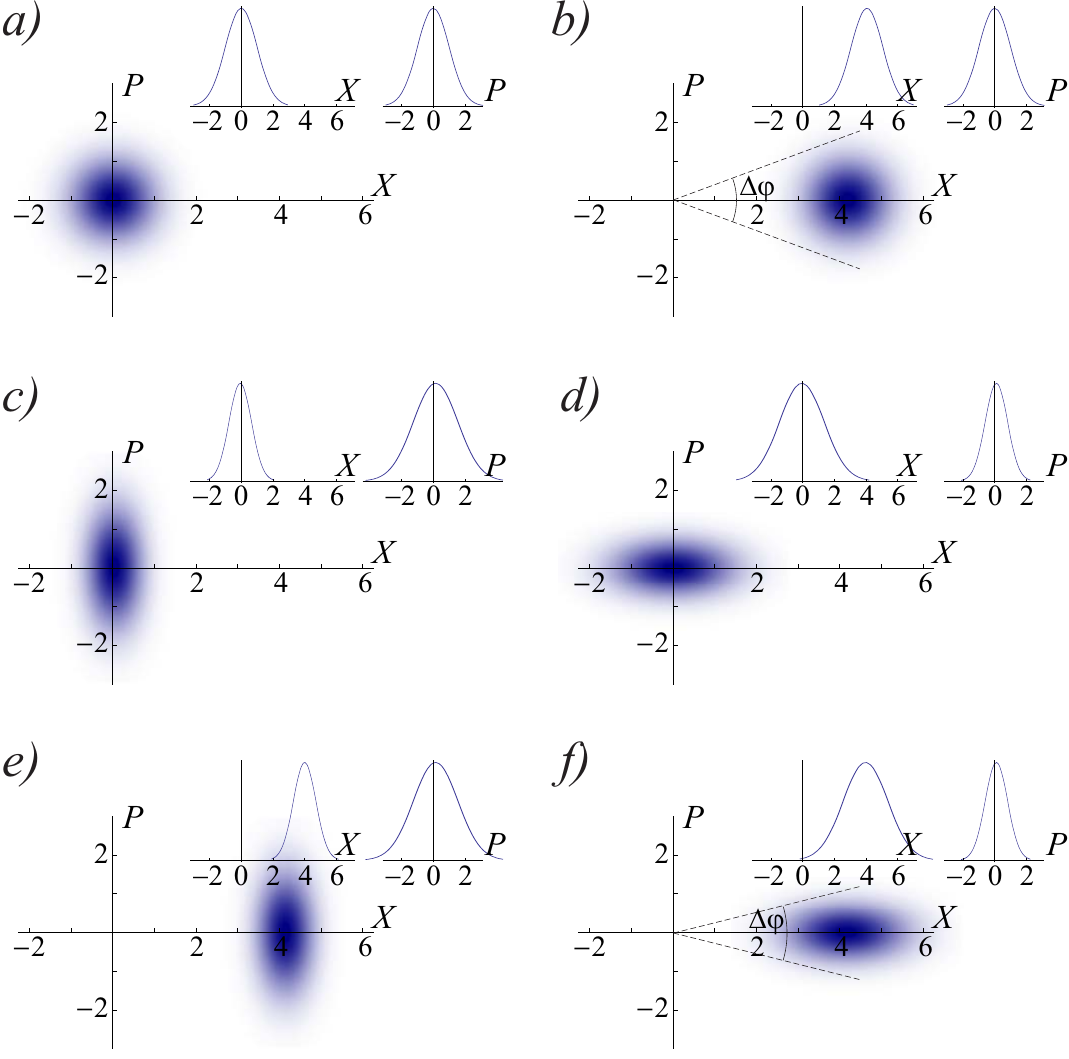}
\caption{Wigner functions of certain single-oscillator states. a) Vacuum state. b) coherent state. c,d) Position- and momentum-squeezed vacuum states. e,f) Position- and momentum-squeezed coherent states with real amplitudes. Panels (b) and (f) show the phase uncertainties of the respective states to emphasize the phase squeezing of state (f). Insets show wavefunctions in the position and momentum bases. \label{SqWigFig}}
\end{figure}

Squeezing is best visualized by means of the Wigner function --- the quantum analogue of the phase-space probability density. Figure \ref{SqWigFig}(c,d) display the Wigner functions of the position- and momentum-squeezed vacuum states, respectively. The squeezing feature becomes apparent when these Wigner functions are compared with that of the vacuum state [Fig.~\ref{SqWigFig}(a)]. Figure \ref{SqWigFig}(e,f) shows \emph{squeezed coherent states}, which are analogous to the squeezed vacuum except that their Wigner function is displaced from the phase space origin akin to the coherent state [Fig. \ref{SqWigFig}(b)].

The state shown in Fig.~\ref{SqWigFig}(f) is particularly interesting because it exhibits, as a consequence of momentum squeezing, \emph{phase squeezing} --- reduction of the uncertainty in the \emph{phase} with respect to a coherent state of the same amplitude. Because the Schr\"odinger evolution under the standard harmonic oscillator Hamiltonian corresponds to clockwise rotation of the phase space around the origin point, the phase squeezing property is preserved under this evolution. In the same context, the state in Fig.~\ref{SqWigFig}(e) is sometimes called \emph{amplitude squeezed}.  

According to the quantum theory of light, the Hilbert space associated with a mode of the electromagnetic field is isomorphic to that of the mechanical harmonic oscillator. The role of the position and momentum observables in this context is played by the electric field magnitudes measured at specific phases. For example, the field at phase zero (with respect to a certain reference) corresponds to the position observable, that at phase $\pi/2$ to the momentum observable, and so on.
Accordingly, phase-sensitive measurements of the field in an electromagnetic wave are affected by quantum uncertainties. For the coherent and vacuum states, this uncertainty is phase-independent  and equals $\sqrt{\hbar\omega/2\varepsilon_0V}$  (the \emph{standard quantum limit}, or \emph{SQL}), where $\omega$ is the optical frequency and $V$ is the quantization volume \cite{LoudonKnight}. But squeezed optical states exhibit uncertainties below SQL at certain phases.

Dependent on whether the mean coherent amplitude of the state is zero, squeezed optical states are classified into squeezed vacuum and \emph{(bright) squeezed light}. Squeezed coherent states form a subset of bright squeezed light states.

How can one generate optical squeezed states in experiment? Consider the state
\begin{equation}\label{zeroplus2}
\ket\psi=\ket 0-\frac s{\sqrt 2} \ket 2,
\end{equation}
where $\ket 0$ and $\ket 2$ are photon number (Fock) states and $s$ is a real positive number. We assume  $s$ to be small, so the norm of state \eqref{zeroplus2} is close to one. The mean value of the position operator $\hat X=(\hat a+\hat a^\dag)/\sqrt 2$ in this state is zero while its variance equals
\begin{equation}\label{zeroplus2var}
\langle \Delta X^2\rangle = \bra\psi\frac{(\hat a+\hat a^\dag)^2}{2}\ket\psi=\frac 12-s,
\end{equation}
so state $\ket\psi$ is position squeezed for positive $s$.

\begin{figure}[h]
\includegraphics[width=\columnwidth]{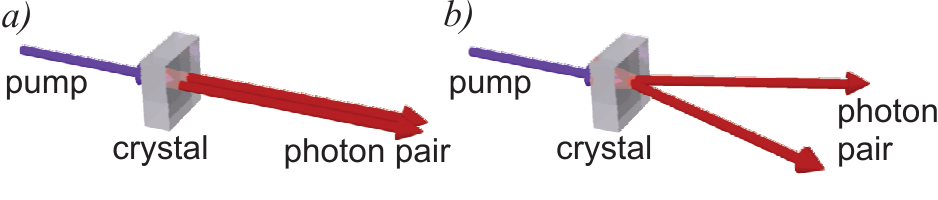}
\caption{Spontaneous parametric down-conversion. a) Degenerate configuration, leading to single-mode squeezed vacuum. b) Non-degenerate configuration, leading to two-mode squeezed vacuum.\label{PDCFig}}
\end{figure}

This result illustrates one of the primary methods of producing squeezing. Spontaneous parametric down-conversion (SPDC) is a nonlinear optical process in which a photon of a powerful laser field propagating through a second-order nonlinear optical medium may split into two photons of lower energy. The frequencies, wavevectors and polarizations of the generated photons are governed by phase-matching conditions. Single-mode squeezing, such as that in the above example, is obtained when SPDC is \emph{degenerate}: the two generated photons are indistinguishable in all their parameters: frequency, direction, and polarization. The quantum state of the optical mode into which the photon pairs are emitted exhibits squeezing [Fig.~\ref{PDCFig}(a)]. 


Aside from being an interesting physical entity by itself, squeezed light has a variety of applications. One of the primary applications of single-mode squeezed light is in precision measurements of distances. Such measurements are typically done by means of interferometry. Quantum phase noise poses an ultimate limit to interferometry, and the application of squeezing (in particular, the phase squeezed state discussed above) permits expanding this limit beyond the fundamental boundary defined by the SQL. For example, squeezing is employed in the new generation of gravitational wave detectors --- GEO 600 in Europe and LIGO in the United States.

\subsection{Two-mode squeezed light}
\begin{figure}[h]
\includegraphics[width=\columnwidth]{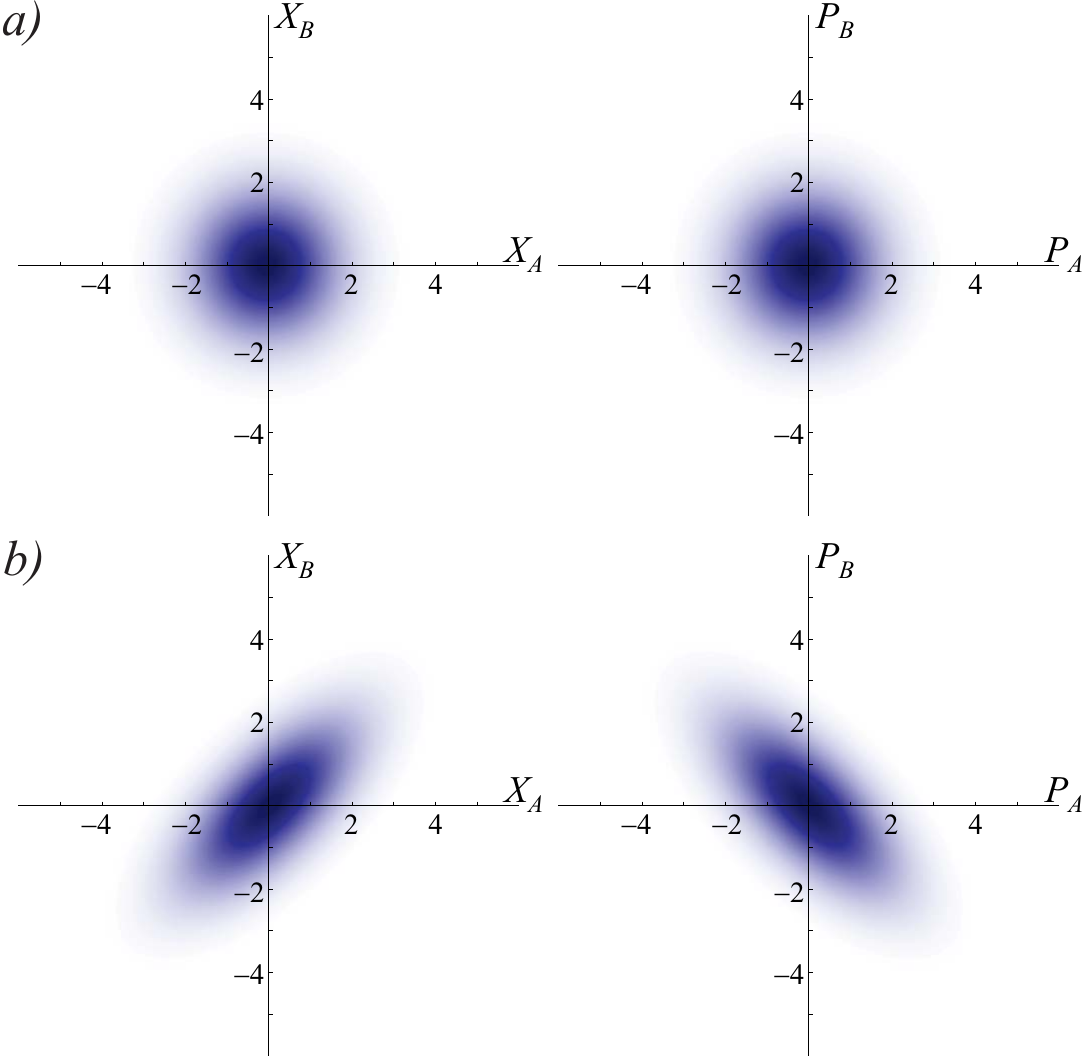}
\caption{\label{TMSVFig}Wavefunctions (not Wigner functions!) of two-mode states in the position (left) and momentum (right) bases. a) The two-mode vacuum state is uncorrelated in both bases. b) The two-mode squeezed state with position observables correlated, and momentum observables anticorrelated beyond the standard quantum limit. }
\end{figure}

A state that is closely related to the single-oscillator squeezed vacuum in its theoretical description and experimental procedures, but quite different in properties is the \emph{two-mode squeezed vacuum} (TMSV), also known as the \emph{twin-beam state}. As the name suggests, this is a state of not one, but two mechanical or electromagnetic oscillators. We introduce this state by first analyzing the tensor product $\ket 0 \otimes\ket 0$ of vacuum states of the two oscillators. In the position basis, its wavefunction [Fig.~\ref{TMSVFig}(a)],
\begin{equation}\label{psi00}
\Psi_{00}(X_a,X_b)=\frac{1}{\sqrt\pi}e^{-X_a^2/2}e^{-X_b^2/2}
\end{equation}
can be rewritten as
\begin{equation}\label{psi00c}
\Psi_{00}(X_a,X_b)=\frac{1}{\sqrt\pi}e^{-(X_a-X_b)^2/4}e^{-(X_a+X_b)^2/4}.
\end{equation}
Here, $X_a$ and $X_b$ are the position observables of the two oscillators which are traditionally associated with fictional experimentalists Alice and Bob. The meaning of \eeqref{psi00c} is that the observables $(X_a-X_b)/\sqrt 2$ and $(X_a+X_b)/\sqrt 2$ have a Gaussian distribution with variance $1/2$. This is not surprising because in the double-vacuum state Alice's and Bob's position observables are uncorrelated and both of them have variance $1/2$. The behavior of the momentum quadratures in this state is analogous to that of the position.

The wavefunction of the two-mode squeezed vacuum state $\tmsv$ is given by
\begin{equation}\label{psiABX}
\Psi_{R}(X_a,X_b)=\frac{1}{\sqrt\pi}e^{-(X_a+X_b)^2/(4R^2)}e^{-R^2(X_a-X_b)^2/4},
\end{equation}
where $R$, as previously, is the squeezing factor [Fig.~\ref{TMSVFig}(c)]. In contrast to the double-vacuum, TMSV is an entangled state, and Alice's and Bob's position observables are nonclassically correlated thanks to that entanglement. For $R>1$, the variance of $(X_a-X_b)/\sqrt 2$ is less than $1/2$, i.e. below the value for the double vacuum state.

The wavefunction of TMSV in the momentum basis is obtained from \eeqref{psiABX} by means of Fourier transform by both Alice's and Bob's observables:
\begin{equation}\label{psiABP}
\tilde\Psi_{R}(P_a,P_b)=\frac{1}{\sqrt\pi}e^{-(P_a-P_b)^2/(4R^2)}e^{-R^2(P_a+P_b)^2/4}.
\end{equation}
We see that for $R>1$ Alice's and Bob's momenta are anticorrelated, i.e. the variance of the sum $(P_a+P_b)/\sqrt 2$ is below the level expected from two vacuum states [Fig.~\ref{TMSVFig}(d)].

The two-mode squeezed vacuum does not imply squeezing in each individual mode. On the contrary,  Alice's and Bob's position and momentum observables in TMSV obey a Gaussian probability distribution with variance
\begin{equation}\label{TMSVind}
\langle\Delta X_a^2\rangle=\langle\Delta X_b^2\rangle=\langle\Delta P_a^2\rangle=\langle\Delta P_b^2\rangle=\frac{1+R^4}{4R^2}.
\end{equation}
that exceeds that of the vacuum state for any $R\ne 1$. In other words, each mode of a TMSV considered individually is in the thermal state. With increasing $R>1$, the uncertainty of individual quadratures increases while that of the difference of Alice's and Bob's position observables as well as the sum of their momentum observables decreases.

In the extreme case of $R\to\infty$, the wavefunctions of the two-modes squeezed state take the form
\begin{eqnarray}
\Psi_{R}(X_a,X_b)&\propto&\delta(X_a-X_b) \label{EPRX}\\
\tilde\Psi_{R}(P_a,P_b)&\propto&\delta(P_a+P_b) \label{EPRP}
\end{eqnarray}
Both Alice's and Bob's positions are completely uncertain, but at the same time precisely equal, whereas the momenta are uncertain but precisely opposite. This state is the basis of the famous quantum nonlocality paradox in its original formulation of Einstein, Podolsky and Rosen (EPR) \cite{EPR}. EPR argued that by choosing to perform either a position or momentum measurement on her portion of the TMSV, Alice remotely prepares either a state with a certain position or one with a certain momentum at Bob's location. But according to the uncertainty principle, certainty of position implies complete uncertainty of momentum, and vice versa. In other words, by choosing the setting of her measurement apparatus, Alice can instantly and remotely, without any interaction, prepare at Bob's station one of two mutually incompatible physical realities. This apparent contradiction to basic principles of causality has lead EPR to challenge quantum mechanics as complete description of physical reality and triggered a debate that continues to this day.

Experimental realization of TMSV is largely similar to that of single-mode squeezing. SPDC is the primary method; however, in contrast to the single-mode case, it is implemented in the non-degenerate configuration. The photons is each generated pair are emitted into two distinguishable modes that become carriers of the TMSV state [Fig.~\ref{PDCFig}(b)].

In order to understand how non-degenerate SPDC leads to squeezing, consider the two-mode state
\begin{equation}\label{zeroplus11}
\ket\Psi=\ket 0\otimes\ket 0+s\ket 1\otimes\ket 1,
\end{equation}
i.e. a pair of photons has been emitted into Alice's and Bob's modes with amplitude $s$. Now if we evaluate the variance of the observable $(X_a-X_b)/\sqrt 2$, we find
\begin{equation}\label{zeroplus11var}
\frac 12\langle \Delta (X_a-X_b)^2\rangle =\frac 14 \bra\Psi(\hat a+\hat a^\dag-\hat b-\hat b^\dag)^2\ket\Psi=\frac 12-s,
\end{equation}
i.e. Alice's and Bob's position observables are correlated akin to TMSV. A similar calculation shows anticorrelation of Alice's and Bob's momentum observables.

Both the single-mode and two-mode squeezed vacuum states are valuable resources in quantum optical information technology. TMSV, in particular, is useful for generating heralded single photons and unconditional quantum teleportation.

\section{Salient features of squeezed states}
\subsection{The squeezing operator}\label{SqOpSec}
We now proceed to a more rigorous mathematical description of squeezing. Single-mode squeezing occurs under the action of operator
\begin{equation}\label{SqOp}
\hat S(\zeta)=\exp[(\zeta\hat a^2-\zeta^*\hat a^{\dag 2})/2],
\end{equation}
where $\zeta=re^{i\phi}$ is the \emph{squeezing parameter}, with $r=\ln R$ and $\phi$ being real numbers, upon the vacuum state. Phase $\phi$ determines the angle of the quadrature that is being squeezed. In the following, we assume this phase to be zero so $\zeta=r$. Note that, for a small $r$, the squeezing operator \eqref{SqOp} acting on the vacuum state, generates state
\begin{equation}\label{zeroplus2a}
\hat S(r)\ket 0\approx [1+(r\hat a^2-r\hat a^{\dag 2})/2]\ket 0 =\ket 0-(r/\sqrt 2) \ket 2,
\end{equation}
which is consistent with  \eeqref{zeroplus2} for $s=r$. 

\begin{figure}[h]
\includegraphics[width=\columnwidth]{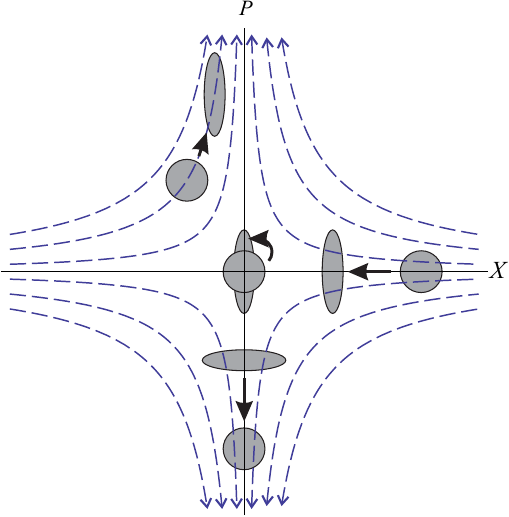}
\caption{Transformation of quadratures under the action of the squeezing Hamiltonian \eqref{SqHam} with $\alpha>0$. Grey areas show examples of Wigner function transformations with $r=\alpha t=\ln 2$.  \label{SqProcessFig}}
\end{figure}

The action of the squeezing operator can be analyzed as fictitious evolution under Hamiltonian
\begin{equation}\label{SqHam}
\hat H=i\hbar\alpha[\hat a^2-(\hat a^\dag)^2]/2
\end{equation}
for time $t=r/\alpha$ (so that $\hat S(r)=e^{-i(\hat H/\hbar) t}$). Analyzing this evolution in the Heisenberg picture, we use $[\hat a, \hat a^\dag]=1$ to find that
\begin{equation}\label{dothataSq}
\dot{\hat{a}} = \frac{i}{\hbar} [\hat{H}, \hat{a}] = -\alpha \hat{a}^\dag
\end{equation}
and
\begin{equation}\label{dotadagSq}
\dot{\hat{a}}^\dag =  -\alpha \hat{a}.
\end{equation}
Now using the expressions for quadrature observables
\begin{equation}\label{quaddefn}
\hat X=(\hat a+\hat a^\dag)/\sqrt 2 \quad{\rm and}\quad \hat P=(\hat a-\hat a^\dag)/\sqrt 2i,
\end{equation}
 we rewrite Eqs.~\eqref{dothataSq} and \eqref{dotadagSq} as
\begin{subequations}
\begin{eqnarray}
\dot{\hat{X}}&=&-\alpha X;\\
\dot{\hat{P}}&=&\alpha P.
\end{eqnarray}
\end{subequations}
If this evolution continues for time $t$, we will have
\begin{subequations}\label{QuadTrafoSq}
\begin{eqnarray}
\hat X(t)&=&\hat S^\dag(r)\hat X(0)\hat S(r)=\hat X(0)e^{-r}; \\
\hat P(t)&=&\hat S^\dag(r)\hat P(0)\hat S(r)=\hat P(0)e^{r},
\end{eqnarray}
\end{subequations}
which corresponds to position squeezing by factor $R=e^r$ and corresponding momentum antisqueezing (Fig.~\ref{SqProcessFig}). If the initial state is vacuum, the evolution will result in a squeezed vacuum state; coherent states will yield squeezed light \cite{Schiller98}.

As a self-check, we find the factor of quadrature squeezing in state \eqref{zeroplus2a}, in analogy to \eeqref{zeroplus2var}:
$$R=\sqrt{\frac{\langle 0 |\Delta X^2| 0 \rangle}{\langle 0 |\hat S^\dag(r)\Delta X^2\hat S(r)| 0 \rangle}}=\sqrt{\frac{1/2}{1/2-r}}\approx 1+r$$ which is in agreement with $R=e^r$ for small $r$.

The corresponding transformation of the creation and annihilation operators is given by
\begin{subequations}\begin{eqnarray}
\hat a(t)&=&\hat a(0)\cosh r-\hat a^\dag(0)\sinh r;\\
\hat a^\dag(t)&=&\hat a^\dag(0)\cosh r-\hat a(0)\sinh r,
\end{eqnarray}\end{subequations}
known as \emph{Bogoliubov transformation}.

Two-mode squeezing is treated similarly. The two-mode squeezing operator is
\begin{equation}\label{SqOp2}
\hat S_2(\zeta)=\exp[(-\zeta\hat a\hat b+\zeta^*\hat a^\dag\hat b^\dag)].
\end{equation}
Assuming, again, a real $\zeta=r$, introducing the fictitious Hamiltonian and recalling that the creation and annihilation operators associated with different modes commute, we find
\begin{subequations}\label{Bogoliubov}\begin{eqnarray}
  \hat a(t)&=& \hat a(0) \cosh r+\hat b(0)^\dag \sinh r; \\
  \hat b(t)&=& \hat b(0) \cosh r+\hat a(0)^\dag \sinh r;
\end{eqnarray}\end{subequations}
and hence
\begin{subequations}\label{BogoliubovXP}\begin{eqnarray}
  \hat X_a(t) \pm \hat X_b(t) &=& [\hat X_a(0) \pm\hat X_b(0)] e^{\pm r}; \\
  \hat P_a(t) \pm \hat P_b(t) &=& [\hat P_a(0) \pm\hat P_b(0)] e^{\mp r}.
\end{eqnarray}\end{subequations}
Initially, Alice's and Bob's modes are  in vacuum states, and the quadrature observables in these modes are uncorrelated. But as the time progresses, Alice's and Bob's position observables become correlated while the momentum observables become anticorrelated.
\subsection{Photon number statistics}
An important component in the theoretical description of squeezed light is its decomposition in the photon number basis, i.e. calculating the quantities $\langle n \sq$ for the single-mode squeezed state and $\langle {mn}\tmsv$ for the two-mode state. Due to non-commutativity of the photon creation and annihilation operators, this calculation turns out surprisingly difficult even for basic squeezed vacuum states, let alone squeezed coherent states and the states that have been affected by losses. Possible approaches to this calculation include the disentangling theorem for SU(1,1) Lie algebra \cite{BuzekKnight}, direct calculation of the wavefunction overlap in the position space \cite{Leonhardt} or transformation of the squeezing operator \cite{Caves85}. Here we derive the photon number statistics of single- and two-mode squeezed vacuum states by calculating their inner product with coherent states.

The wavefunction of a coherent state with real amplitude $\alpha$ is
\begin{equation}\label{cohwf}
\psi_\alpha(X)=\frac{1}{\pi^{1/4}}e^{-(X-\alpha\sqrt 2)^2/2},
\end{equation}
so its inner product with the position squeezed state \eqref{wfsqX} equals
\begin{equation}\label{Ralpha}
\langle \alpha \sq=\intinf\psi_\alpha(X)\psi_R(X)\de X=\sqrt\frac{2R}{1+R^2}e^{-\frac{R^2}{1+R^2}\alpha^2}.
\end{equation}
Now we recall that the coherent state is decomposed into the Fock basis according to
\begin{equation}\label{cohFock}
\ket\alpha=\sum\limits_{n=0}^\infty e^{-\alpha^2/2}\frac{\alpha^n}{\sqrt {n!}}\ket n,
\end{equation}
so we have
\begin{equation}\label{nsq1}
\sum\limits_{n=0}^\infty\langle n \sq \frac{\alpha^n}{\sqrt {n!}} =\sqrt\frac{2R}{1+R^2}e^{\frac{1-R^2}{2(1+R^2)}\alpha^2}
\end{equation}
Decomposing the exponent in right-hand side of the above equation into the Taylor series with respect to $\alpha$, we obtain
\begin{equation}\label{nsq2}
\sum\limits_{n=0}^\infty\langle n \sq \frac{\alpha^n}{\sqrt n!} =\sqrt\frac{2R}{1+R^2}\sum\limits_{m=0}^\infty
\left[\frac{1-R^2}{2(1+R^2)}\right]^m\frac{\alpha^{2m}}{m!}.
\end{equation}
Because this equality must hold for any real $\alpha$, each term of the sum in the left-hand side must equal its counterpart in the right-hand side that contains the same power of $\alpha$. Hence $n=2m$ and
\begin{equation}\label{nsq3}
\langle {2m} \sq =\sqrt\frac{2R}{1+R^2}\left[\frac{1-R^2}{2(1+R^2)}\right]^m\frac{\sqrt{(2m)!}}{m!}.
\end{equation}
Since $R=e^r$, we have
\begin{equation}\label{RtoHyp}
\frac{2R}{1+R^2}=\frac 1{\cosh r}\quad\textrm{and}\quad\frac{1-R^2}{1+R^2}=-\tanh r,
\end{equation}
so \eeqref{nsq3} can be rewritten as
\begin{equation}\label{nsq4}
\sq =\frac 1{\sqrt{\cosh r}}\sum\limits_{m=0}^\infty(-\tanh r)^m\frac{\sqrt{(2m)!}}{2^m m!}\ket{2m}.
\end{equation}

We stop here for a brief discussion. First, we note that that for $r\ll 1$, \eeqref{nsq4} becomes
\begin{equation}\label{nsq4exp}
\sq=\ket 0-(r/\sqrt 2) \ket 2+O(r^2),
\end{equation}
consistently with \eeqref{zeroplus2a}. Second, note that the squeezed vacuum state \eqref{nsq4} contains only terms with even photon numbers. This is a fundamental feature of this state; in fact, one of the earlier names for squeezed states has been ``two-photon coherent states" \cite{Yuen}. This feature follows from the nature of the squeezing operator \eqref{SqOp}: in its decomposition into the Taylor series with respect to $r$, creation and annihilation operators occur only in pairs. Pairwise emission of photons is also a part of the physical nature of SPDC: due to energy conservation a pump photon can only split into two photons of half its energy.

We now turn to finding the photon number decomposition of the two-mode squeezed state. We first notice, by looking at \eeqref{SqOp2}, that $\ket{R_{AB}}$ must only contain terms with equal photon numbers in Alice's and Bob's modes. This circumstance allows us to significantly simplify the algebra. We proceed along the same route as outlined above, calculating the overlap of $\ket{{\rm  TMSV}_R}$ with the tensor product $\ket{\alpha\alpha}$ of identical coherent states $\ket\alpha$ in Alice's and Bob's channels using Eqs.~\eqref{psiABX} and \eqref{cohwf}:
\begin{eqnarray}
\langle\alpha\alpha{\tmsv}&&\nonumber\\
&&\hspace{-2cm}=\intinf\psi_\alpha(X_a)\psi_\alpha(X_b)\Psi_{R}(X_a,X_b)\de X_a\de X_b\nonumber\\ &&\hspace{-2cm}=\frac{2R}{1+R^2}e^{-\frac{2}{1+R^2}\alpha^2}.\label{RABalpha}
\end{eqnarray}
Decomposing the coherent states in the left-hand side into the Fock basis according to \eeqref{cohFock} and keeping only the terms with equal photon numbers, we have
\begin{equation}\label{n2sq}
\sum\limits_{n=0}^\infty\braket {nn}{{\rm  TMSV}_R} \frac{\alpha^{2n}}{n!} =\frac{2R}{1+R^2}e^{-\frac{1-R^2}{1+R^2}\alpha^2}
\end{equation}
Now writing the Taylor series for the right-rand side and using \eeqref{RtoHyp}, we obtain
\begin{equation}\label{n2sq4}
\tmsv  =\sum\limits_{n=0}^\infty\frac{1}{\cosh r}\tanh^n r\ket{nn}.
\end{equation}

Similarly to the single-mode squeezing, it is easy to verify that the above result is consistent with state \eqref{zeroplus11} for small $r$. On the other hand, in contrast to the single-mode case, the energy spectrum of TMSV follows Boltzmann distribution with mean photon number in each mode $\langle n\rangle=\sinh^2 r$. This is in agreement with our earlier observation that Alice's and Bob's portions of TMSV  considered independently of their counterpart are in the thermal state, i.e. the state whose photon number distribution obeys Boltzmann statistics with the temperature given by $e^{-\hbar\omega/kT}=\tanh r$.

 While the present analysis is limited to pure squeezed vacuum states, photon number decompositions of squeezed coherent states and squeezed states that have undergone losses can be found in the literature \cite{Scully,Dodonov94}. In contrast to pure squeezed vacuum states, these decompositions have nonzero terms associated to non-paired photons. The origin of these terms is easily understood. If a one- or two-mode squeezed vacuum state experiences a loss, it may happen that one of the photons in a pair is lost while the other one remains. If the squeezing operator acts on a coherent state, the odd photon number terms will appear in the resulting state because they are present initially.

Photon statistics of both classes of squeezed states have been tested experimentally, as discussed in Section \ref{BHDSec} below. An example is shown in Fig.~\ref{BreitenbachPhotonsFig}.

\begin{figure}[h]
\includegraphics[width=0.7\columnwidth]{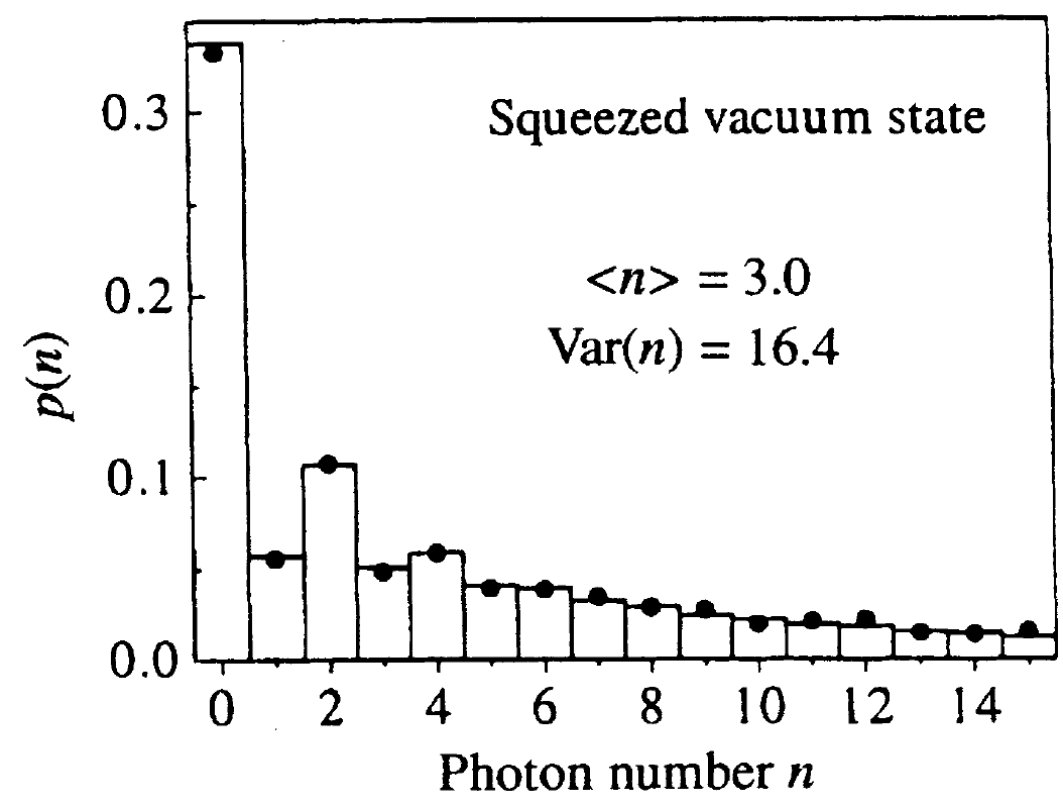}
\caption{Experimentally reconstructed photon number statistics of the squeezed vacuum state. For low photon numbers, the even terms are greater than the odd terms due to pairwise production of photons, albeit the odd term contribution is nonzero due to loss. Reproduced from Ref.~\cite{Breitenbach97}. \label{BreitenbachPhotonsFig}}
\end{figure}


\subsection{Interconversion between single- and two-mode squeezing}\label{InterconvSec}
\begin{figure}[h]
\includegraphics[width=\columnwidth]{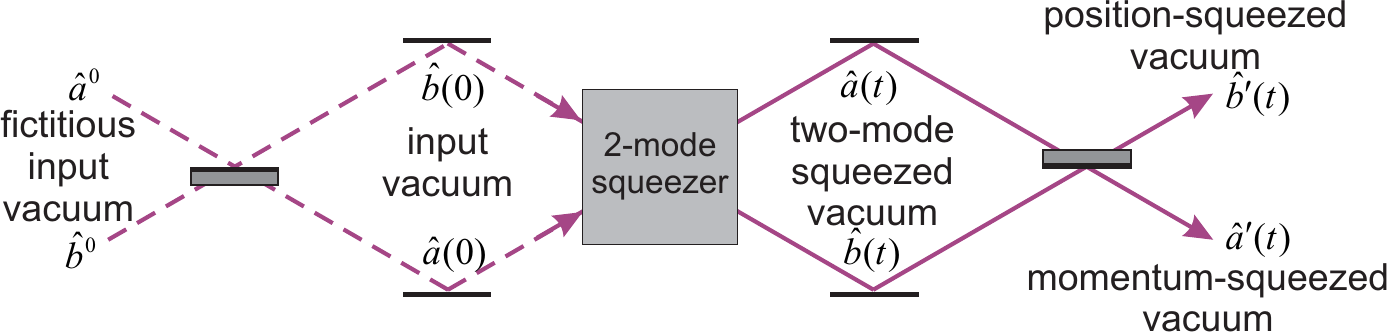}
\caption{\label{InterconvFig}Interconversion of the two-mode squeezed vacuum and two single-mode squeezed vacuum states. Dashed lines show a fictitious beam splitter transformation of a pair of vacuum states such that the modes $\hat a'(t),\hat b'(t)$ are explicitly single-mode squeezed with respect to modes $\hat a'^0,\hat b'^0$.}
\end{figure}

If the modes of the TMSV are overlapped on a symmetric beam splitter, two unentangled single-mode vacuum states will emerge in the output  (Fig.~\ref{InterconvFig}). To see this, we recall the beam splitter transformation
\begin{subequations}\label{BSOpTrans}
\begin{eqnarray}
\hat a' &=&   \tau\hat a-\rho\hat b; \\
\hat b' &=&   \tau\hat b+\rho\hat a,
\end{eqnarray}
\end{subequations}
where $\tau$ and $\rho$ are the beam splitter amplitude transmissivity and reflectivity, respectively. For a symmetric beam splitter, $\tau=\rho=1/\sqrt 2$. In writing Eqs.~\eqref{BSOpTrans}, we neglected possible phase shifts that may be applied to individual input and output modes \cite{Leonhardt}.

In accordance with the definition \eqref{quaddefn} of quadrature observables, Eqs.~\eqref{BSOpTrans} apply in the same way to the position and momentum of the input and output modes. Applying this to Eqs.~\eqref{BogoliubovXP}, we find \begin{eqnarray}\label{BSOpEPRX}
\hat X'_{a,b} &=&   [\hat X_a(t)\mp\hat X_b(t)]/\sqrt2 \nna
&=& e^{\mp r}[\hat X_a(0)\mp\hat X_b(0)]/\sqrt2
\end{eqnarray}
for the output positions and
\begin{eqnarray}\label{BSOpEPRP}
\hat P'_{a,b} &=&   [\hat P_a(t)\mp\hat P_b(t)]/\sqrt2 \nna
&=& e^{\pm r}[\hat P_a(0)\mp\hat P_b(0)]/\sqrt2
\end{eqnarray}
for the momenta. In order to understand what state this corresponds to, let us assume, for the sake of the argument, that vacuum modes $\hat a$ and $\hat b$ at the SPDC input have been obtained from another pair of modes by means of another symmetric beam splitter:
\begin{eqnarray}\label{BSOpInit}
\hat a^0 &=&  [\hat a(0)-\hat b(0)]/\sqrt2\\
\hat b^0 &=&  [\hat a(0)+\hat b(0)]/\sqrt2.
\end{eqnarray}
Of course, since modes $\hat a(0)$ and $\hat b(0)$ are in the vacuum state, so are $\hat a^0$ and $\hat b^0$. We then have:
\begin{eqnarray}\label{BSOpInitQuad}
\hat X'_{a,b} &=&   e^{\mp r}\hat X^0_{a,b};\nna
\hat P'_{a,b} &=&   e^{\pm r}\hat P^0_{a,b},
\end{eqnarray}
where superscript $0$ associates the quadrature with modes $\hat a^0$ and $\hat b^0$. We see that modes $\hat a'$ and $\hat b'$ are related to vacuum modes $\hat a^0$ and $\hat b^0$ by means of position and momentum squeezing transformations, respectively.

Because the beam-splitter transformation is reversible, it can also be used to obtain a TMSV from two single-mode squeezed vacuum states with squeezing in orthogonal quadratures. This technique has been used, for example, in the experiment on continuous-variable quantum teleportation \cite{KimbleQT}.
\subsection{Squeezed vacuum and squeezed light}\label{DispSec}
Squeezed vacuum and bright squeezed light are readily converted between each other by means of the phase-space displacement operator \cite{Leonhardt}, whose action in the Heisenberg picture can be written as
\begin{equation}\label{DispOpAct}
\hat D^\dag(\alpha)\hat a^\dag\hat D(\alpha)=\hat a^\dag+\alpha.
\end{equation}
This means, in particular, that the position and momentum transform according to
\begin{eqnarray}
\hat X&\mapsto&\hat X+{\rm Re}\,\alpha\sqrt 2;\\
\hat P&\mapsto&\hat P+{\rm Im}\,\alpha\sqrt 2,
\end{eqnarray}
so, under the action of $\hat D(\alpha)$, the entire phase space displaces itself, thereby changing the coherent amplitude of the squeezed state without changing the degree of squeezing.

\begin{figure}[h]
\includegraphics[width=0.6\columnwidth]{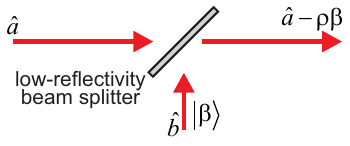}
\caption{\label{DispFig}Implementation of phase-space displacement. $\rho\ll 1$ is the beam splitter's amplitude reflectivity.}
\end{figure}

Phase-space displacement can be implemented experimentally by overlapping the signal state with a strong coherent state $\ket\beta$ on a low-reflectivity beam splitter (Fig.~\ref{DispFig}). Applying the beam splitter transformation \eqref{BSOpTrans}, we find for the signal mode
\begin{equation}\label{DispExp}
\hat a' =   \tau\hat a-\rho\hat b
\end{equation}
Given that mode $\hat b$ is in a coherent state (i.e. an eignestate of $\hat b$) and that $\rho\ll 1$ (i.e. $\tau\sim 1$), we have
\begin{equation}\label{DispExp1}
\hat a' =   \hat a-\rho \beta
\end{equation}
in analogy to \eeqref{DispOpAct}. The displacement operation has been used to change the amplitude of squeezed light in many experiments, for example, in Ref.~\cite{Schnabel04}.

\subsection{Effect of losses}
\begin{figure}[h]
\includegraphics[width=0.6\columnwidth]{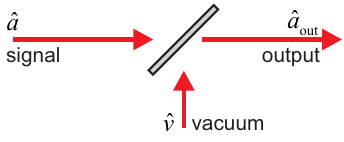}
\caption{\label{BSModelFig}The beam splitter model of loss.}
\end{figure}

Squeezed states that occur in practical experiments necessarily suffer from losses present in sources, transmission channels and detectors. In order to understand the effect of propagation losses on a single-mode squeezed vacuum state, we can use the model in which a lossy optical element with transmission $T$ is replaced by a beam splitter (Fig.~\ref{BSModelFig}). At the other input port of the beam splitter there is a vacuum state. The interference of the signal mode $\hat a$ with the vacuum mode $\hat v$ will produce a mode with operator $\hat a'=\tau\hat a-\rho\hat v$ (with $\tau^2=T$ and $\rho^2=1-T$ being the beam splitter transmissivity and reflectivity) in the beam splitter output. Accordingly, we have
\begin{equation}\label{BSModel}
\hat X_{\theta,{\rm out}}=\tau\hat X_{a,\theta}-\rho\hat X_{v,\theta} .
\end{equation}
Because the quadrature observable of the signal and vacuum states are uncorrelated, and since $\avg{\Delta (X_{v,\theta})^2}=1/2$, it follows that
\begin{eqnarray}\label{BSModelUnc}
\avg{\Delta X_{\theta,{\rm out}}^2} &=&   \tau^2\avg{\Delta (X_{a,\theta})^2}+\rho^2\avg{\Delta (X_{v,\theta})^2}\nna
&=&T\avg{\Delta (X_{a,\theta})^2}+(1-T)/2.
\end{eqnarray}
Analyzing Eqs.~\eqref{BSOpTrans} we see that the optical loss alone, no matter how significant, cannot eliminate the property of squeezing completely.

Ideal squeezed-vacuum and coherent states have the minimum-uncertainty property: the product of uncertainties $\avg{\Delta X_{\rm out}^2}\avg{\Delta P_{\rm out}^2}$ reaches the theoretical minimum of $1/4$. But this is no longer the case in the presence of losses. The deviation of the uncertainty from the minimum  can be used to estimate the preparation quality of a squeezed state. Suppose a measurement of a squeezed state yielded the minimum and maximum quadrature uncertainty values of $\avg{\Delta X_{\theta,{\rm out}}^2}$ corresponding to $\theta=0$ and $\theta=\pi/2$, respectively. One can assume that the state has been obtained from an ideal (minimum-uncertainty) squeezed state with squeezing $R$ by means of loss channel with transmissivity $T$. Using \eeqref{UncSimple}, one can write Eq.~\eqref{BSModelUnc} for $\theta=0$ and $\theta=\pi/2$ and solve the obtained system of equations for $T$ and $R$. These values can then be compared with those expected from the setup at hand in order to  find out if any unexpected losses are present in it \cite{Wasilewski06}.

\section{Detection}\label{BHDSec}
\subsection{Balanced homodyne detection}
In order to detect squeezing, we need to perform multiple measurements of the field quadrature, i.e. the observable $\hat X_\theta=\hat X\cos\theta+\hat P\sin\theta=[e^{-i\theta}\hat A+e^{i\theta}\hat A^\dag]/\sqrt 2$, where $\hat A$ is the annihilation operator of the mode of interest. The task of measuring optical fields in a phase-sensitive fashion may appear daunting, as these fields oscillate at frequencies on a scale of hundreds of terahertz. Fortunately however, such a measurement can be implemented using a relatively simple interference setup. The technique known as balanced homodyne detection, proposed in 1983 by Yuen and Chan \cite{Yuen83} and subsequently implemented by Abbas \ieac \cite{Abbas83} to this day remains the method of choice for quadrature measurements. Reference \cite{LvovskyRaymer} provides a review of the current state of the art in this area.

Here I start with a brief overview of this technique, in the way it is presented in most textbooks. Subsequently, I will discuss a more complex but important question of identifying the temporal mode whose quadrature is being measured. For simplicity, I will start in the classical language.
\begin{figure}[h]
\includegraphics[width=\columnwidth]{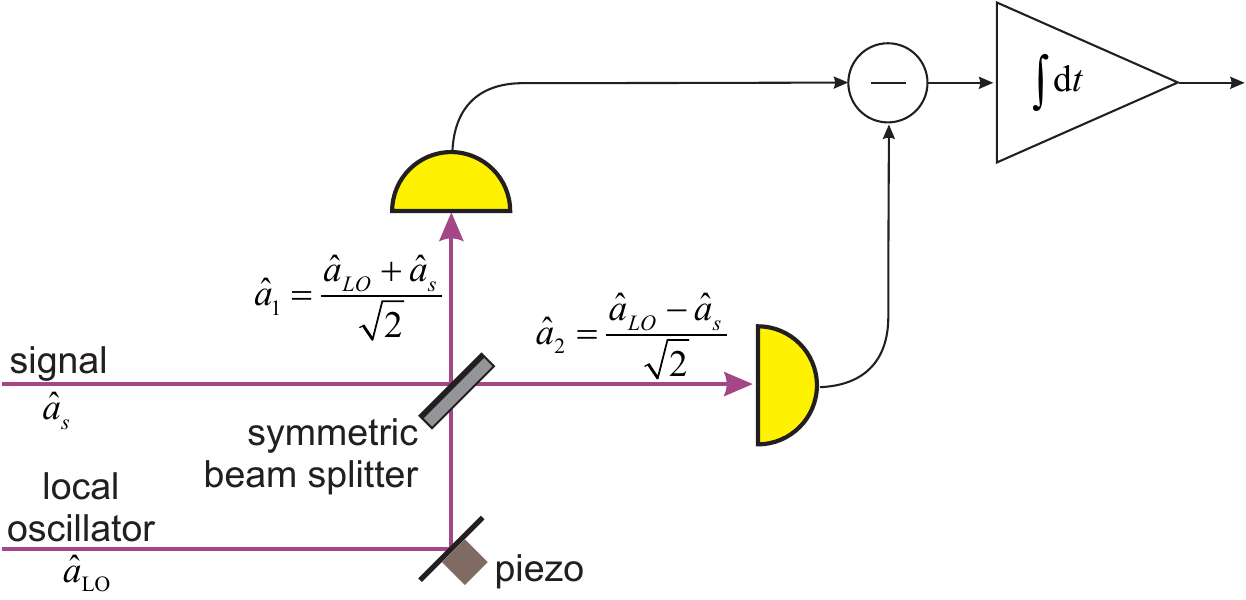}
\caption{\label{HDSchemeFig}The principle of balanced homodyne detection.}
\end{figure}

Suppose the field to be measured (referred to as \emph{signal}) is centered at frequency $\Omega$. We write for the field magnitude $E(t)\propto a(t)e^{-i\Omega t}+a^*(t)e^{i\Omega t}$, where $a(t)$ is slowly varying. For quadrature measurement, this field is overlapped on a symmetric (50:50) beam splitter with a strong laser field at frequency $\Omega$, known as the \emph{local oscillator (LO)}, with amplitude $a_{\rm LO}(t)$: $E_{\rm LO}(t)\propto a_{\rm LO}(t)e^{-i\Omega t+i\theta}+a_{\rm LO}^*(t)e^{i\Omega t-i\theta}$. The phase $\theta$ of the local oscillator is controlled, e.g. by a piezoelectric transducer. The two beam splitter output fields impinge onto two photodiodes whose output photocurrents are electronically subtracted (Fig.~\ref{HDSchemeFig}).

In order to see how the subtracted photocurrent in the detector output relates to the signal quadrature, we write the amplitudes of the beam splitter outputs as
\begin{equation}\label{OHTBS}
a_{1,2}(t)=\frac{a_{\rm LO}(t)e^{i\theta}\pm a(t)}{\sqrt 2}.
\end{equation}
The detector's output current is then proportional to the intensity difference
\begin{eqnarray}\label{Nminus}
I_-(t)&\propto& a_1(t)^*a_1(t)-a_2(t)^*a_2(t)\nna&=&a(t)a^*_{\rm LO}(t)e^{-i\theta}+a^*(t)a_{\rm LO}(t)e^{i\theta}.
\end{eqnarray}
Assuming that $a_{\rm LO}(t)$ is real (this is a matter of choosing the zero phase reference point), the quantity in the right-hand side of \eeqref{Nminus} is an instantaneous (or, rather, averaged over the detection electronics' response time) value of the classical quadrature $a(t)e^{-i\theta}+a^*(t)e^{i\theta}$.


Switching to quantum treatment, we replace the classical amplitude $a(t)$ by operator $\hat a(t)$. This operator is defined as
\begin{equation}\label{at}
\hat a(t)=\frac 1{\sqrt{2\pi}}\intinf\hat a_\omega e^{-i(\omega-\Omega) t}\de \omega,
\end{equation}
where $\hat a_\omega$ is the annihilation operator of a plane wave mode of optical frequency $\omega$ familiar from the electromagnetic field quantization procedure. One can think of  $\hat a(t)$ as the annihilation operator of a photon of frequency $\Omega$ at time moment $t$. Such a description is of course unphysical because of the time-frequency uncertainty principle; yet sometimes it turns out useful for visualization.

As to the local oscillator, we recall that it is in a high-amplitude coherent state so the relative quantum noise of its amplitude is negligible. Hence we can continue to treat the LO amplitude $a_{\rm LO}(t)$ as a number, not an operator. Equation \eqref{Nminus} simplifies to
\begin{equation}\label{Iminus}
\hat I_-(t)\propto \alpha_{\rm LO}(t)[\hat a(t)e^{-i\theta}+\hat a^\dag(t)e^{i\theta}].
\end{equation}


There are two primary approaches to the acquisition and analysis of the subtraction photocurrent of the homodyne detector. In \emph{time-domain} analysis, the photocurrent is measured using a time-resolving device, such as an oscilloscope. In \emph{frequency-domain} measurements, one instead looks at the electronic spectrum of the photocurrent.

\subsection{Time-domain approach}
In the time-domain approach, the goal is to measure the quadrature of a limited duration temporal mode defined by annihilation operator
\begin{equation}\label{tempmode}
\hat A=\intinf \varphi(t)\hat a(t)\de t,
\end{equation}
where $\varphi(\cdot)$ is some normalized real function of bounded support. As follows from \eeqref{Iminus}, this measurement can be realized by multiplying the subtraction photocurrent, obtained from the homodyne detector with a constant LO, by the mode function and integrating it over time:
\begin{equation}\label{tempmeas}
\intinf  \varphi(t)\hat I(t)\de t\propto a_{\rm LO}(\hat A e^{-i\theta}+\hat A^\dag e^{i\theta})=\sqrt 2 a_{\rm LO} \hat X_\theta.
\end{equation}
This approach works if the temporal resolution of the acquisition electronics (typically on a scale of nanoseconds) is fast compared to the duration of the mode of interest, such as in Ref.~\cite{Wasilewski09}.

The opposite extreme that frequently occurs in experimental practice is that the squeezed state is prepared using a pico- or femtosecond pulsed laser, and its temporal mode is defined by the laser pulse. In this case, the quadrature measurement can be accomplished in spite of lack of resolution at the electronic level by using the same laser as the local oscillator. We then have $a_{\rm LO}(t)\propto\varphi(t)$ and hence $\intinf \hat I(t)\de t\propto  \hat X_\theta$. Because of the slow electronics' response, the integration occurs in this setting automatically. The output of the homodyne detector is an electrical pulse whose shape is determined by the response function, and magnitude is proportional to the quadrature \cite{Hansen01}. 


Time-domain homodyne detection permits full reconstruction of the state in the acquisition mode. By varying the local oscillator phase $\theta$, one can obtain noise statistics for all quadratures. Probability distributions $\pr(X_\theta)=\braketop{X_\theta}{\hat\rho}{X_\theta}$ for all phase angles are sufficient to obtain full information about the density operator $\hat\rho$ of the signal state, such as its Wigner function or the photon number representation. This method of measuring the quantum state of light is referred to as \emph{optical homodyne tomography} \cite{Leonhardt,LvovskyRaymer}.

Homodyne tomography was first proposed in 1989 \cite{VogelRisken} and implemented experimentally in application to single-mode squeezed vacuum in 1993 \cite{Smithey93} and to two-mode squeezed vacuum in 2000 \cite{Vasilyev00}.

\subsection{Frequency-domain approach}
\begin{figure}[h]
\includegraphics[width=\columnwidth]{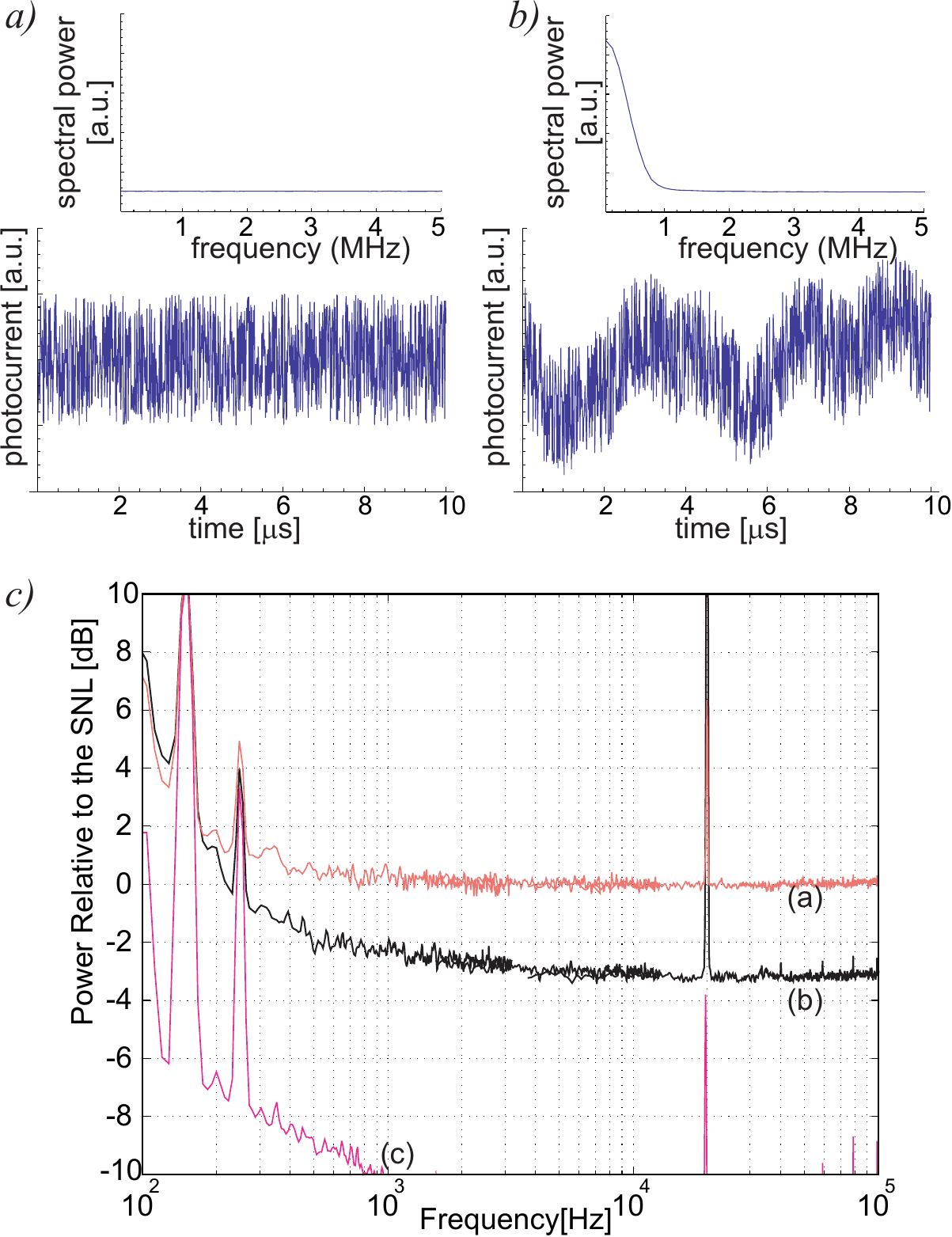}
\caption{\label{SpectrumPrincipleFig} Frequency-domain approach to homodyne measurements. a) Simulated output of a homodyne detector exhibiting noise corresponding to the SQL (bottom). Its spectrum (top) is flat. b) The same output affected by the drift of the zero point on a time scale of 2$\mu$s. Direct variance measurement of the photocurrent will not give SQL. However, the spectral power remains at the SQL level for frequencies above 1 MHz and hence allows observation of squeezing. c) \emph{Bona fide} spectrum of the homodyne detector from Ref.~\cite{McKenzie2004} showing similar behavior. Curve $a$ corresponds to SQL, $b$ to squeezed vacuum and $c$ to the detector's electronic noise (in the absence of LO). The noise peaks arise from harmonics of 50 Hz power line and the phase locking signal (20 kHz).}
\end{figure}

Theoretically, if squeezing is generated in a continuous nonlinear process, it could be observed by measuring the variance of the homodyne detector output photocurrent as a function of the local oscillator phase. In practice, however, this measurement is obscured by various spurious noises produced by either the source or the detector. For example, the reflectivity of the homodyne detector's beam splitter can vary as a function of time due to minute perturbations to its orientation. However small such variation may be, it may affect precise subtraction of the LO amplitudes. As a result, the mean value of the output photocurrent will drift with time, and the drift amplitude can exceed the shot noise level, thereby obscuring the observation of quantum noise (Fig.~\ref{SpectrumPrincipleFig}).

Fortunately, such technical (classical) noises of the photocurrent can be distinguished from the quantum noise by analyzing their spectral behavior. Technical noises often occur within specific frequency bands; for example, the slow drift of the zero point is limited to low frequencies [Fig.~\ref{SpectrumPrincipleFig}(b)]. The quantum noise, on the other hand, is ``white": it is constant for all frequencies within the detector's bandwidth [Fig.~\ref{SpectrumPrincipleFig}(a)]. One can therefore observe squeezing, even in the presence of technical noises, by only looking at those sidebands in which they do not appear [Fig.~\ref{SpectrumPrincipleFig}(c)]. A further advantage of the frequency-domain method is that, by measuring the quantum noise at different sidebands, one is able to analyze the properties of the source and detector; in particular, measure the spectral band in which the squeezing is present.

We start our theoretical analysis of frequency-domain measurements by find the Fourier transform of the photocurrent \eqref{Iminus}  using \eeqref{at}:
\begin{equation}\label{Inu}
\tilde{\hat I}(\nu)=\frac 1{\sqrt{2\pi}}\intinf \hat I(t)e^{i\nu t}\de t\propto\hat a_{\Omega+\nu}e^{-i\theta}+\hat a^\dag_{\Omega-\nu}e^{i\theta},
\end{equation}
ehre $\nu$ is the electronic frequency. By some algebra, we can express the right-hand side of this equation as
\begin{eqnarray}\label{Inu1}
&&\hat a_{\Omega+\nu}e^{-i\theta}+\hat a^\dag_{\Omega-\nu}e^{i\theta}\nnb
&&\hspace{0.5cm}=\hat X_+\cos\theta+\hat P_+\sin\theta-i\hat X_-\sin\theta+i\hat P_-\cos\theta\nnb
&&\hspace{0.5cm}=\hat X_{+,\theta}+i\hat X_{-,\pi/2+\theta}
\end{eqnarray}
where
\begin{equation}\label{XPpmfreq}
\hat X_{\pm,\theta}=\frac 1{\sqrt2}[\hat X_{\theta}(\Omega+\nu)\pm\hat X_{\theta}(\Omega-\nu)].
\end{equation}
This means that measuring the real and imaginary parts of $\tilde{\hat I}(\nu)$ is equivalent to subjecting frequency modes $\Omega+\nu$ and $\Omega-\nu$ to a beam splitter operation and performing homodyne measurements of the beam splitter outputs at LO phases $\theta$ and $\pi/2+\theta$ \cite{Barbosa13}.

Suppose that the state entering the homodyne detector is squeezed --- that is, the noise of $\hat I(t)$ is below the standard quantum limit for a certain local oscillator phase $\theta$. But this would also imply that its Fourier transform --- both the real and imaginary parts of $\tilde{\hat I}(\nu)$ --- exhibit fluctuations below SQL. This would in turn mean that observables $\hat X_{+,\theta}$ and $\hat X_{-,\pi/2+\theta}$ exhibit reduced variance at the same time --- that is, modes  $\Omega+\nu$ and $\Omega-\nu$ are in the two-mode squeezed state (\emph{cf.} Sec.~\ref{InterconvSec}). In other words, \emph{single-mode squeezing in the time domain is equivalent to two-mode squeezing in the frequency domain}. An explicit experimental demonstration to that effect has been presented by Huntington \iea \cite{Huntington05}.

\begin{figure}[h]
\includegraphics[width=\columnwidth]{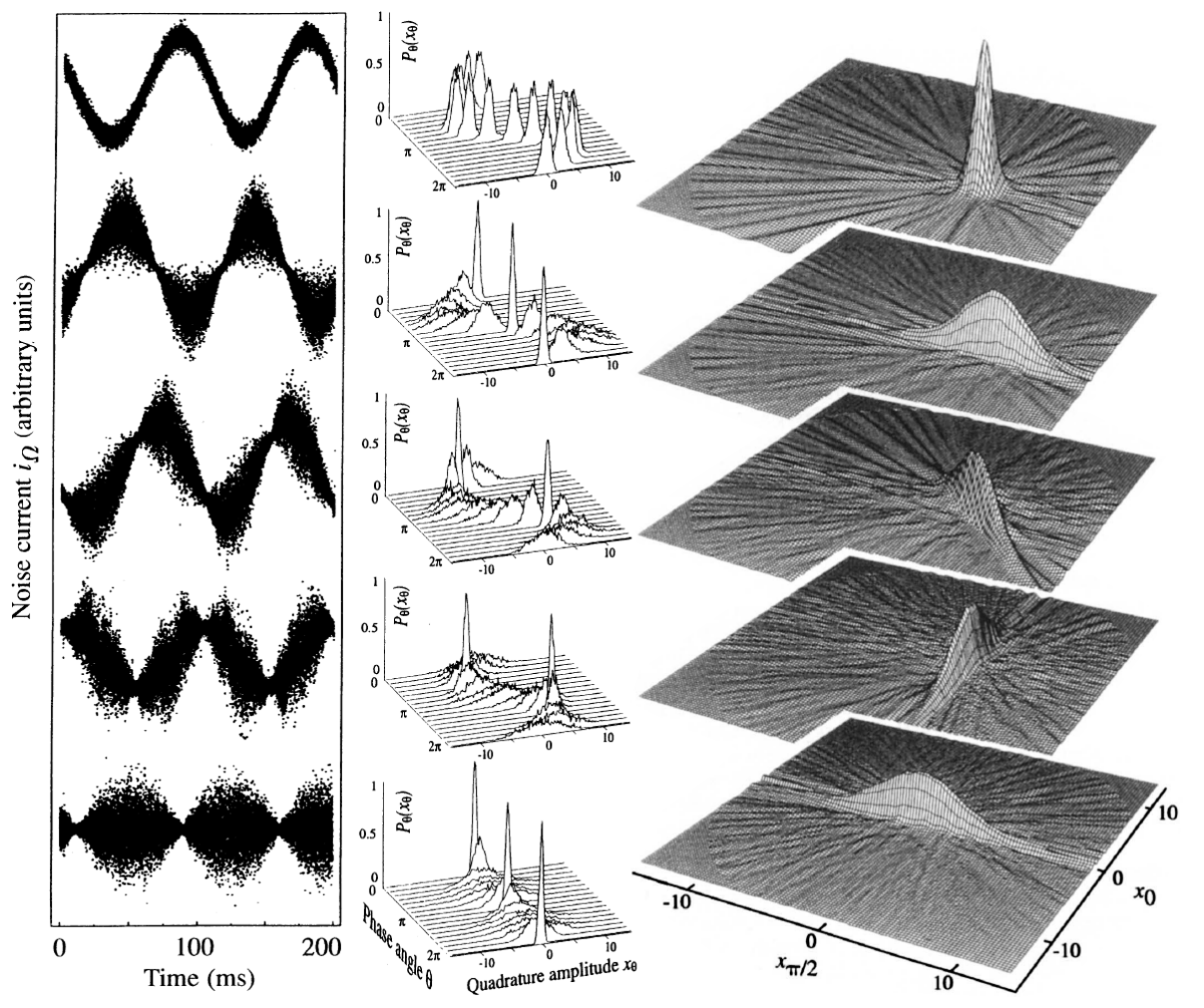}
\caption{\label{BreitenbachFig} Optical homodyne measurements of coherent and squeezed states of light. Top to bottom: coherent state, phase (momentum) squeezed light, $45^\circ$ quadrature squeezed light, amplitude (position) squeezed light, momentum squeezed vacuum. Left column: statistics of quadrature measurements obtained with a balanced homodyne detector while the local oscillator phase is varied. Middle column: histograms $\pr(X_\theta)$ of these measurements associated with specific values $\theta$ of the phase. These histograms are integral projections of the Wigner functions onto vertical planes positioned at angle $\theta$ with respect to the position axis of the phase space. The histograms are used to reconstruct the Wigner fucntions (right column) of the corresponding states in a procedure similar to computer tomography scanning in medicine. They were also used to reconstruct the states' density matrices in the photon number basis (Fig.~\ref{BreitenbachPhotonsFig}) by means of the quantum state sampling method \cite{Leonhardt}. Reproduced from Ref.~\cite{Breitenbach97}.}
\end{figure}

Simultaneous measurements of the real and imaginary parts of $\tilde{\hat I}(\nu)$ are possible using lock-in amplifiers. In this way, one can perform full quantum-state tomography of the modes defined by operators $(\hat a_{\Omega+\nu}\pm\hat a^\dag_{\Omega-\nu})/\sqrt 2$. In a classic work of 1997, Breitanbach \iea \cite{Breitenbach97} used this approach for tomography of an extended family of Gaussian states, including coherent, squeezed vacuum, as well as amplitude and phase squeezed light states (Fig.~\ref{BreitenbachFig}).

It is common to use the electronic spectrum analyzer rather than lock-in amplifiers for frequency-domain measurements. The spectrum analyzer displays the  mean squared power of the photocurrent's sideband:
\begin{equation}\label{spectran}
\avg{\left|\tilde{\hat I}(\nu)\right|^2}\propto\avg{\hat X_{+,\theta}^2+\hat X_{-,\pi/2+\theta}^2}.
\end{equation}
Because observables   $\hat X_{+,\theta}$ and $\hat X_{-,\pi/2+\theta}$ are simultaneously squeezed, the spectrum analyzer will show reduced signal at frequency $\nu$. In this way, the spectrum analyzer can measure squeezing in spite of being unable to resolve the two terms of \eeqref{Inu1}. This result is consistent with the common sense expectation: if the time-dependent photocurrent $I(t)$ exhibits reduced noise, so will its frequency spectrum.

An important shortcoming of the spectrum analyzer is that it does not enable quantum state tomography. It provides information about the variance of the quadrature probability distribution, but not the probability distribution itself. This does not matter, however, if the variance (i.e. the amount of squeezing) is the only quantity of interest, and state reconstruction is not the goal.

\section{Preparation}
In Sec.~\ref{SqOpSec} we had a conceptual discussion of a Hamiltonian that squeezes the phase space. But in fact, almost any Hamiltonian that is at least quadratic in the creation and annihilation operators brings about sophisticated trajectories in the phase space and can result in squeezing. Similarly, a Hamiltonian that is bilinear in the creation and annihilation operators of two modes is likely to generate two-mode squeezing. Accordingly, there exist many physical processes that can be employed to prepare single- and two-mode squeezed states of the electromagnetic field.


An important limitation to the above is the requirement that the Hamiltonian evolution that leads to squeezing be not compromised by competing non-unitary processes that increase noise. For example, attempts to achieve squeezing in atomic systems have for a long time met with limited success due to incoherent spontaneous emission into the signal mode, which leads to thermalization of the signal state and loss of squeezing.

Most frequently, squeezing is obtained by nonlinear optical wave mixing processes, in which pairs of photons are emitted into degenerate (single-mode squeezing) or non-degenerate (two-mode squeezing) modes. An example is spontaneous parametric down-conversion, a three-wave mixing between the pump field and the two photons of squeezed vacuum that occurs due to second-order optical nonlinearity. A related method is four-wave mixing, a third-order nonlinear process in which two strong waves, interacting with a nonlinear medium, give rise to a pair of photons. Let us discuss these two processes in more detail.

\subsection{Via parametric down-conversion}
In order to mathematically describe nonlinear-optical squeezing, we begin with equations for the propagation of classical electromagnetic fields through a nonlinear medium. We then quantize the fields and replace their amplitudes with corresponding creation and annihilation operators, thereby obtaining their evolution in the Heisenberg picture. 

Consider a three-wave mixing process in which a strong \emph{pump} field of frequency $2\Omega$ interacts with weak \emph{signal} and \emph{idler} fields of frequencies $\Omega\pm\nu$, respectively, with $\nu\ll\Omega$, in a crystal with effective nonlinearity $\chi_{\rm eff}$. All fields are continuous in time, but the amplitudes\footnote{The amplitude is defined according to $E(z,t)=\varepsilon(z)e^{ikz-i\omega t}+{\rm c.c.}$, for $E(z,t)$ being the value of the field in space in time.} $\varepsilon_s(z)$ and $\varepsilon_s(z)$ of the signal and idler change with the propagation distance $z$ due to the nonlinear interaction. The pump amplitude $\varepsilon_p$ is assumed to remain constant because $\varepsilon_p\gg \varepsilon_s,\varepsilon_i$, so there is no depletion. We further assume that the crystal is perfectly phase matched for this nonlinear process.

In the slowly-varying envelope approximation \cite{Boyd}, the equations of motion for the signal and idler fields take the form
\begin{equation}\label{SVEANL}
\frac\partial{\partial_z}\varepsilon_{s,i}(z)=i\frac{(\Omega\pm\nu)}{2\epsilon_0nc}P_{\rm NL}(\Omega\pm\nu),
\end{equation}
where $n$ the refractive index and the nonlinear polarization amplitude is given by
\begin{equation}\label{PNL}
P_{\rm NL}(\Omega\pm\nu)=2\varepsilon_0\chi_{\rm eff}\varepsilon_p\varepsilon_{i,s}^*(z),
\end{equation}
$\chi_{\rm eff}$ being the effective nonlinear succeptibility.
Without loss of generality, we can define the phase of the pump such that $ia_p$ is real and positive. Then, solving these equations for propagation length $L$ under assumption $\nu\ll\Omega$, we find
\begin{subequations}\label{CWESoln}
\begin{eqnarray}
  \varepsilon_s(L) &=& \varepsilon_s(0)\cosh r+\varepsilon^*_i(0)\sinh r;\\
  \varepsilon_i(L) &=& \varepsilon_i(0)\cosh r+\varepsilon^*_s(0)\sinh r
\end{eqnarray}
\end{subequations}
with
\begin{equation}\label{rNLO}
r=\frac{\chi_{\rm eff}\Omega}{n c} |a_p|L.
\end{equation}
We now quantize the signal and idler field according to $\varepsilon_{s,i}\to \sqrt{\hbar\Omega/2\epsilon_0 V}\hat a_{\Omega\pm\nu}$ (where $V$ is the quantization volume), but continue to treat the macroscopic pump field as classical. This leads to
\begin{equation}\label{BogoPDC}
  \hat a_{\Omega\pm\nu}(L) = \hat a_{\Omega\pm\nu}(0)\cosh r+\hat a^\dag_{\Omega\mp\nu}(0)\sinh r;\\
\end{equation}
which is identical to Eqs.~\eqref{Bogoliubov}. In other words, if the signal and idler fields of frequencies $\Omega\pm\nu$ before the crystal are in the vacuum state, they will be in a two-mode squeezed state after the crystal.

As discussed in the previous section, such a state manifests itself as single-mode squeezing when a homodyne measurement with the local oscillator tuned to frequency $\Omega$ is performed. To see this, consider a time-domain mode whose annihilation operator is given by
\eeqref{tempmode}. Using \eeqref{at}, we rewrite the mode operator as
\begin{eqnarray}\label{tempmodefreq}
\hat A&=&\intinf\hat a_\omega\tilde\varphi(\omega-\Omega)\de \omega\nna
&=&\int\limits_0^{+\infty}[\hat a_{\Omega+\nu}\tilde\varphi(\nu)+\hat a_{\Omega-\nu}\tilde\varphi(-\nu)]\de\nu,
\end{eqnarray}
where $\tilde\varphi(\nu)=(1/\sqrt{2\pi})\intinf\varphi(t)e^{i\nu t}\de t$ is the Fourier image of $\varphi(t)$. Because $\varphi(t)$ is real, we have $\tilde\varphi(\nu)=\tilde\varphi^*(-\nu)$. Using \eeqref{BogoPDC}, we find
\begin{eqnarray}
  \hat A(L) &=& \int\limits_0^{+\infty}\cosh r[\hat a_{\Omega+\nu}(0)\tilde\varphi^*(\nu)+\hat a_{\Omega-\nu}(0)\tilde\varphi(\nu)]\nna
  &&\hspace{2mm}+\sinh r[\hat a^\dag_{\Omega+\nu}(0)\tilde\varphi(\nu)+\hat a^\dag_{\Omega-\nu}(0)\tilde\varphi^*(\nu)]\de\nu\nna
  &=&\hat A(0)\cosh r+\hat A^\dag(0)\sinh r,
\end{eqnarray}
i.e. single-mode squeezing. The above derivation is valid only if $\varphi(t)$ is sufficiently slowly varying so that its spectrum $\tilde\varphi(\nu)$ takes on significant values only at such frequencies  that two-mode squeezing of operators $\hat a_{\Omega\pm\nu}$ is present. In practice, this limitation is established by the nonlinear crystal's phase-matching bandwidth (for single-pass squeezing) or the cavity linewidth (for squeezing in a cavity).

One can readily estimate the amount of squeezing one can obtain. Consider, for example,  a $L=5$ mm periodically poled KTP crystal with the signal wavelength of $\lambda=780$ nm and the pump field of power $P=100$ mW focused into a spot of $w=50\ \mu$m radius. The relevant effective nonlinear coefficient of PPKTP is $\chi_{\rm eff}=14$ pm/V, refractive index $n=1.8$.

Under these conditions, the pump intensity is $I_p=P/\pi w^2=1.3\times 10^7$ W/m$^2$ and the field amplitude $|\varepsilon_p|=\sqrt{I_p/2n\epsilon_0c}=3.6\times 10^4$ V/m. Substituting this value into \eeqref{rNLO}, we find $r=1.1\times 10^{-2}$. We see that the amount of squeezing obtained by a single pass of a continuous-wave pump laser through a nonlinear crystal of a reasonable size is very small.

There are two primary methods of addressing this complication. First, one could use an ultrashort pulsed laser, thereby greatly increasing the pump amplitude. The above theory, developed for continuous-wave pump, has only limited application for pulsed pump; the amount of squeezing strongly depends on the shape $\varphi(t)$ of the temporal mode chosen for the measurement \cite{Wasilewski06}. Nevertheless, squeezing has been demonstrated in the single-pass pulsed regime as soon as one year after the first experimental observation of squeezed light \cite{Slusher87} and the degree of squeezing has been increased to several decibels\footnote{Decibel [dB] is a common unit of squeezing in experiment. The degree of squeezing in decibels is calculated according to $10\log_{10}(2\avg{\Delta X^2})$. The standard quantum limit corresponds to a squeezing of 0 dB, the reduction of quadrature variance by a factor of 2 to about 3 dB,  factor of 4 to about 6 dB,  factor of 10 to 10 dB, etc.}
 in subsequent years \cite{Kim94}.

The second approach is to place the crystal inside a Fabry-Perot cavity. The cavity can be resonant to the pump light, thereby enhancing the effective pump power, or to the signal, effectively allowing multiple passing of the signal through the crystal, or both. The case when the cavity is resonant to the signal is most common; this configuration is referred to as the \emph{optical parametric oscillator or amplifier (OPO/OPA)}. A theory of squeezing inside an OPA has been developed by Gardiner and Savage \cite{Gardiner84} and reviewed, for example, in \cite{Scully}. Without derivation, we present the
result for the quadrature noise levels associated with the antisqueezed ($\theta^+=\pi/2$) and squeezed  ($\theta^-=0$) quadratures:
\begin{equation}\label{OPAout}
V^{\pm}(\nu)=\frac 1 2 \pm\eta\frac{2\sqrt{P/P_{th}}}{(\nu/\gamma)^2+(1\mp\sqrt{P/P_{th}})^2},
\end{equation}
where $2\gamma$ is the cavity linewidth, $\eta$ is the overall quantum efficiency, $P$ is the pump power and $P_{th}$ is the threshold power, i.e. the pump power at which the nonlinear process in the cavity leads to macroscopic optical oscillations. By analyzing this result, we see that the squeezing occurs at sideband frequencies $\nu$ less than or on the order of the cavity linewidth. This is not surprising: the enhancement effect of the cavity is only present within its resonance.

\begin{figure}[h]
\includegraphics[width=0.7\columnwidth]{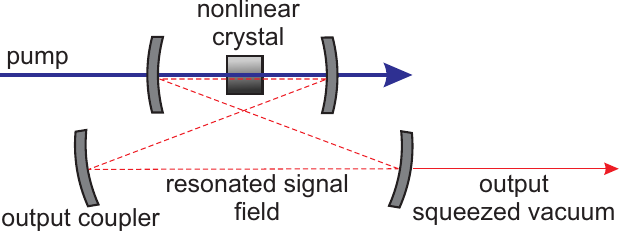}
\caption{\label{bowtieFig} Squeezing in an OPA cavity. The cavity mirrors are reflective to the signal field, but transparent to the pump.}
\end{figure}

The squeezing is strongest at the threshold point, when the amplification in a single pass through the nonlinear crystal is equal to the loss occurring in a roundtrip of the signal through the cavity, including that at the output coupling mirror. The intensity gain factor equals $e^{2r}\approx 1+2r$ for $r\ll1$. In the numerical example studied above, $2r=0.022$, so in order to be at the threshold, the cavity must have the same roundtrip loss. This loss occurs due to the transmission through the output coupling mirror as well as  spurious losses on all other optical elements inside the cavity. Assuming, for example, that the mirror has a transmissivity of $0.017$, and the spurious losses add up to $0.005$, we find for the quantum efficiency $\eta=0.015/(0.017+0.005)=0.77$, which means that at the threshold, for $\nu\ll\gamma$, we will see a variance of $V^-\approx 1/2-\eta/2$, or about 6 dB.

Let us now estimate the bandwidth within which the squeezing is generated. This bandwidth is the same as the cavity linewidth $\gamma$, which, in turn, is the ratio of the cavity's free spectral range and finesse. Assuming that the cavity is of a bow-tie configuration (Fig.~\ref{bowtieFig}) with a full length of $L_c=30$ cm, its free spectral range is $c/L=1$ GHz. The finesse is $\pi/T\approx 160$, so $\gamma\approx 6 MHz$.

Historically, the first observation of squeezing using an OPA cavity has been achieved by Wu \iea in 1986 \cite{Wu86}. The squeezing reached in that experiment was about 3 decibels. Since then, many groups made efforts to further develop this approach. One of the most recent results reported a squeezing of 12.7 dB \cite{Eberle10}. This remarkable achievement required the overall quantum efficiency (including that of the OPA cavity, homodyne detection, mode matching, etc.) to approach 95\%.

OPAs can as well be used successfully to generate two-mode squeezing. The first experiment to that effect was reported by Ou \iea in 1992 \cite{Ou92}. In that work, the signal and idler fields  resonated in the cavity were of the same frequency, but different polarizations.


\subsection{In atomic ensembles}
As mentioned above, high optical nonlinearity is at the heart of most squeezing processes. An atom interacting with an optical wave resonant with one of its transitions is an intrinsically nonlinear object. Atoms begin to exhibit nonlinear optical properties at intensity levels on a scale of the saturation intensity, which is many orders of magnitude lower than the intensity levels required for significant nonlinear effects in ferroelectric crystals. Therefore atomic ensembles have been considered an attractive medium for the preparation of squeezed optical states from early days of quantum optics.

\begin{figure}[h]
\includegraphics[width=0.5\columnwidth]{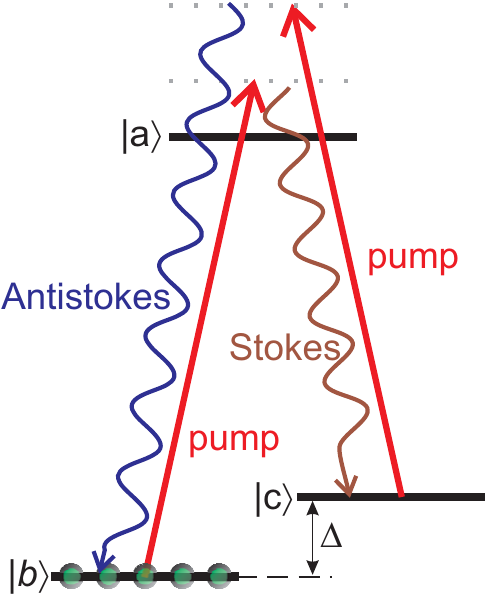}
\caption{\label{FWMFig} Quantum four-wave mixing in an atomic $\Lambda$ system leads to emission of the Stokes and anti-Stokes photons akin to signal and idler in parametric down conversion. Two-mode squeezing obtains for non-degenerate and single-mode squeezing for degenerate ground states. }
\end{figure}

A typical mechanism that leads to the generation of squeezing is four-wave mixing (Fig.~\ref{FWMFig}). Consider a $\Lambda$-shaped atomic energy level configuration with two ground states coupled to a single excited state by optical transitions of degenerate or nondegenerate frequencies. This configuration is present, in particular, in alkali atoms, where the ground level is split into two hyperfine sublevels.

Suppose the atom is initially in ground state $\ket b$. The pump field of frequency $\Omega$ excites the $\ket a-\ket b$ transition, driving the atom into the other ground state $\ket c$ through Raman scattering, which results in emission of a Stokes photon of frequency $\Omega-\Delta$. Level $\ket c$ is in turn excited by the pump field and the atom goes back into $\ket b$, accompanied by emission of an anti-Stokes photon  of frequency $\Omega+\Delta$. If the entire process is coherent, the Stokes and anti-Stokes emission modes will find themselves in the two-mode squeezed state.

This mechanism was used in the very first observation of optical squeezing by Slusher \iea \cite{Slusher85}.  In that experiment, atomic vapor of sodium has been used and two Fabry-Perot cavities resonant with the pump and the Stokes/Antistokes fields have been placed around the vapor sample for amplification. A two-mode squeezed state at frequencies $\Omega\pm\Delta$ was observed using a homodyne detector with the local oscillator at frequency $\Omega$. A squeezing of about 0.3 dB has been detected for correlated quadratures, while the uncorrelated quadratures exhibited extra noise at a level of about 2 dB.

The state observed by Slusher and co-workers did not approach the minimum-uncertainty limit. This is largely due to processes in atoms that occur concurrently to four-wave mixing and lead to incoherent emission into the signal modes, such as Brillouin and Raman scattering. Further hindrance is presented by various dephasing phenomena such as time-of-flight decoherence that inhibit coherent four-wave mixing. All these processes contribute to the ``thermalization" of the optical state in the signal modes and degrade the squeezing.

This appears to be a common problem in experiments using atomic ensembles for squeezing. This is the primary reason that ferroelectric crystals, rather than atomic systems became the workhorse of squeezed light generation. In recent years, however, atomic systems have been revisited and significant squeezing has been demonstrated in experiments involving four-wave mixing \cite{Lett07,Lett08} and polarization self-rotation \cite{Ries03, Lezama11}.
\subsection{In fibers}
Optical fibers are typically made of glass, an amorphous material with inversion symmetry. Accordingly, they normally possess no second-order nonlinearity. However, fibers enable propagation of focused optical wavepackets over long distances, so the effects of third-order (Kerr) nonlinearities on these wavepackets become significant. One of these effects is squeezing.

\begin{figure}[h]
\includegraphics[width=0.7\columnwidth]{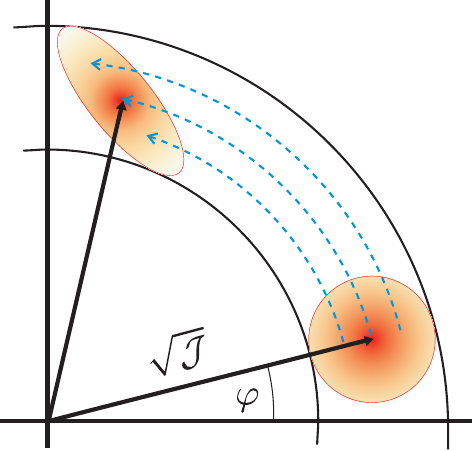}
\caption{\label{LeuchsFig} Effect of a Kerr medium on a coherent state. Different intensities experience different refractive indices, resulting in quadrature squeezing. From Ref.~\cite{Rigas13}}
\end{figure}

Squeezing in optical fiber is best explained in terms of the nonlinear refractive index. In a Kerr medium, the refractive index depends on intensity $I$ of the propagating light according to
\begin{equation}\label{NLn}
n=n_0+n_2I,
\end{equation}
where $n_2$ is related to the third-order nonlinear susceptibility $\chi^{(3)}$. The phase of light that has propagated through such a material will then depend on the intensity, resulting in the transformation of the Wigner function as illustrated in Fig.~\ref{LeuchsFig}. The parts of the Wigner function that are associated with higher and lower intensities becomes shifted in the phase space with respect to each other, resulting in squeezing.

Homodyne detection of squeezing in this configuration is complicated by the macroscopic mean amplitude of the signal required to take advantage of \eeqref{NLn}. The amplitude could, in principle, be eliminated by means of phase-space displacement (see Sec.~\ref{DispSec}); however, this would require a powerful laser and excellent phase stabilization. A more common detection method involves causing two fiber squeezed fields to interfere with each other so the resulting phase-space displacement and rotation makes one of the resulting fields amplitude squeezed. Amplitude squeezing is then readily observed by measuring the intensity with a single high-efficiency detector and evaluating the variance of the photocurrent noise.

There are a number of ways such interference can be implemented. For example, in a Sagnac-type interferometer the initial laser pulse impinges on a beamsplitter, after which the transmitted and reflected fields enter the fiber from two ends. Upon exiting the fiber, the fields interfere on the same beam splitter, and one of the resulting fields is measured \cite{Drummond93}. Alternatively, a polarization-maintaining fiber is used, so that the fields in both polarizations become squeezed at the fiber output. These fields are then brought into interference using waveplates positioned at the output end of the fiber \cite{Heersink05}.

Squeezing in optical fibers is limited by phase noise associated with thermal fluctuations of the refractive index, in particular guided acoustic wave Brilloun scattering. An additional degrading factor, particularly significant for very short pulses, is Raman scattering \cite{Drummond93}. Both these phenomena allow precise theoretical treatment, and can be minimized by wise choice of experimental parameters \cite{Corney06,Dong08}. As a result, squeezing up to about 7 dB has been obtained \cite{Dong08}.

\section{Applications in quantum information}
Squeezed light is a primary resource in continuous-variable\footnote{The term `continuous-variable' refers to optical quantum information protocols that involve manipulation of a state in phase space, i.e. displacement, squeezing, quadrature measurements, etc. It is usually contrasted with `discrete-variable' methods dealing with manipulating and measuring single photons. This separation is largely of historical and technological nature; in fact, more and more interesting applications now arise at the boundary between the two domains \cite{LvovskyRaymer,AndersenReview}.} quantum information processing. In addition to fundamental interest such as the implementation of the original EPR paradox, it is the basis of many basic applications such as universal quantum computing, dense coding and quantum key distribution. The limited volume of this manuscript does not permit a comprehensive review of these applications; such a review can be found, for example, in Refs.~\cite{Braunstein05,Andersen10}. Here we will concentrate on only two important examples.

\subsection{Quantum-optical state engineering}
Lasers generate coherent states and their statistical mixtures --- the states of light known as classical. While such states are useful for some applications, many emerging quantum technologies require a supply of optical states that lie outside the classical domain. Nonclassical optical states cannot be achieved by linear-optical manipulation: interference of coherent states necessarily leads to coherent states. Production of nonclassical states therefore requires nonlinear optics.

Parametric down-conversion is a nonlinear phenomenon capable of producing quantum states of light with high efficiencies and with well-defined spatiotemporal properties. This property is unique among existing methods of non-classical light generation (see, e.g. Ref.~\cite{SinglePhotonsOnDemand}). However, the only states that SPDC can produce are  the single- and dual-mode squeezed vacua. For this reason, the past decade has seen extensive efforts to use these states as ``primitives" to produce (``engineer") various other states of light. As we see in this section, application of tools such as linear-optical manipulation, interference with coherent states and conditional measurements allows one to accomplish this task successfully. However small the degree of squeezing may be, even a single squeezed resource permits producing a wide variety of complex optical states \cite{LvovskyRaymer,Illuminati06}.

A TMSV with a weak level of squeezing can be used to generate \emph{heralded single photons}. To that end, one channel of that state (idler) is monitored by a single-photon detector. If the detector ``clicks", we know, according to \eeqref{n2sq4}, that a photon must have been emitted into the other (signal) channel as well. If the squeezing parameter $r$ is sufficiently small, the contribution of higher photon numbers in the signal channel can be neglected.

In 2001, this technique was used to generate a heralded single photon in a definite spatiotemporal mode, characterize it using homodyne tomography and, for the first time, observe a negative Wigner function ~\cite{Lvovsky01}. This method was later extended to generate and measure the two-  \cite{Ourjoumtsev06a} and three-photon \cite{Cooper13,Yukawa13} states. In these extensions, the idler channel of SPDC was split into multiple photon detectors, and their coincident ``clicks" were required for a heralding event.

\begin{figure}
 \includegraphics[width=\columnwidth]{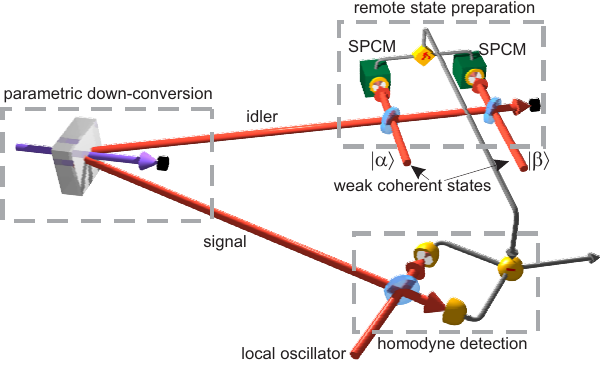}
\caption{\label{TwoPhotonEngineeringFig}Generating arbitrary superpositions of the zero-, one- and two-photon states. The light in the idler channel of parametric down-conversion is brought into interference with two weak coherent states and subsequently detected by single-photon counting modules (SPCMs). A double ``click" heralds the generation of the desired state in the signal channel.}
\end{figure}

A modification of this scheme shown in Fig.~\ref{TwoPhotonEngineeringFig} permits producing arbitrary superpositions of photon number states. Prior to detection, the light in the idler channel is mixed with weak ancillary coherent states on beam splitters. In this way, a detector registering a photon ``does not know" whether it comes from SPDC or from a coherent state. This indistinguishability results in the idler channel of SPDC being effectively projected onto a superposition of Fock states. Thanks to entanglement of the TMSV, this superposition is automatically transferred to the signal channel. The weight of each component of the superposition can be controlled by the amplitudes and phases of the ancilla coherent states. This technique has been demonstrated for superpositions of up to the two- \cite{Bimbard10} and three-photon \cite{Yukawa13} terms, but can, in principle, be extended to higher numbers. One of the possible applications of this method is the implementation of the cubic phase gate for universal quantum computation in the continuous-variable setting.

In the above examples,  a low magnitude of the squeezing parameter does not degrade the fidelity of engineered quantum states. Quite the contrary, it ensures that the state is not contaminated by higher photon number components. However, low squeezing also reduces the probability of the heralding event, which can make the method unpractical. One must choose the degree of squeezing as a compromise between the fidelity and the state production rate.

In the next example, in contrast, a non-negligible value of squeezing is essential for obtaining the desired state --- a superposition $\ket\alpha\pm\ket{-\alpha}$ of two coherent states of opposite amplitudes. This state is  of interest to the quantum community because, while being  a linear combination of classical states, it is highly nonclassical,  and hence reminiscent of the famous ``Schr\"odinger cat" Gedankenexperiment.

Remarkably, the squeezed vacuum is quite similar to the state $\ket\alpha+\ket{-\alpha}$ (``even Schr\"odinger kitten") for  $\alpha\lesssim 1$. To see this, recall the Fock decomposition \eqref{cohFock} of the coherent state. The sum of two coherent states of opposite amplitudes will contain only even photon number terms,
\begin{equation}\label{evencat}
\ket\alpha+\ket{-\alpha}\propto\ket 0+\frac{\alpha^2}{\sqrt 2}\ket 2+O(\alpha^4),
\end{equation}
in the same way as the squeezed vacuum state \eqref{nsq4exp}. With a sufficiently small $\alpha$, only the first two terms of these decompositions are significant, and setting $r=\alpha^2$ makes them mutually identical for the two states.

\begin{figure}
\includegraphics[keepaspectratio,width=\columnwidth]{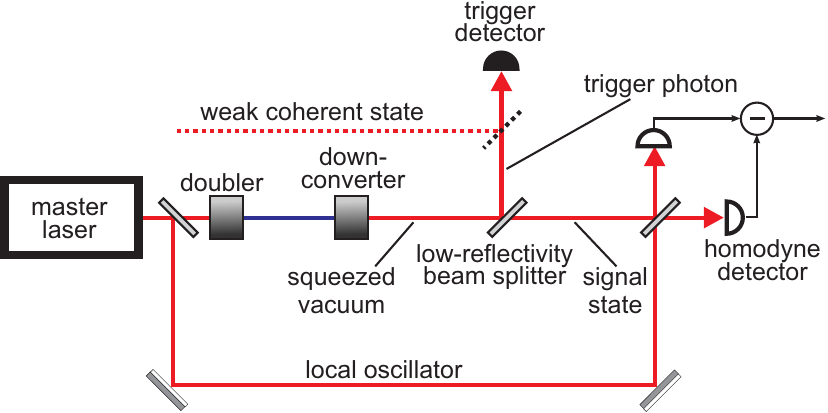}
\vspace*{8pt}
\caption{\label{kittenFig}(a) Conditional preparation of the ``odd Schr\"odinger kitten'' by applying photon annihilation to the squeezed vacuum state (the ``even Schr\"odinger kitten''). The dashed lines shows the additional elements used to generate arbitrary superpositions of states $\ket\alpha$ and $\ket{-\alpha}$ as in Ref.~\cite{Neergaard10}.}
\end{figure}

Because coherent states are eigenstates of the photon annihilation operator $\hat a$, applying that operator to $\ket\alpha+\ket{-\alpha}$ produces  $\alpha(\ket\alpha-\ket{-\alpha})$, i.e. an ``odd Schr\"odinger kitten".  This idea was implemented experimentally by Wenger \iea \cite{Wenger04} and later refined in Refs.~\cite{Ourjoumtsev06,Neergaard06,Wakui07,Gerrits10}. For photon annihilation, squeezed vacuum produced by means of degenerate SPDC was transmitted through a low-reflectivity beam splitter (Fig.~\ref{kittenFig}). Detection of a photon in the reflected channel indicates that a photon has been removed from the squeezed vacuum --- that is, a photon annihilation event has occurred \cite{Kumar13}.

If a weak ancilla coherent state is injected into the heralding detector using an additional beam splitter (Fig.~\ref{kittenFig}), the heralding photon cannot be definitively traced back to the squeezed state or that ancilla. If the event comes from the squeezed state, the photon subtraction takes place and the signal output is the odd Schr\"odinger kitten; if it comes from the coherent state, the signal output is the same as the input, i.e. the even Schr\"odinger kitten. Because these two possibilities are indistinguishable, the output state becomes a coherent superposition of the even and odd kittens, with the magnitude and phase of the terms in the superposition dependent on the parameters of the ancilla. In this way, arbitrary superpositions of states $\ket\alpha$ and $\ket{-\alpha}$ --- an optical continuous-variable qubit --- are generated \cite{Neergaard10}.

\subsection{Continuous-variable quantum teleportation}
Teleportation is a quantum communication protocol in which a quantum state is transferred between two locations without utilizing any direct quantum communication channel. The transfer is enacted by local interference of the signal state with a portion of the entangled resource shared between the two locations, as well as local measurements, classical communications and local quantum operations. The teleportation protocol was first proposed for qubits in 1993 by Bennett \iea \cite{Bennett93}, and for continuous variables in 1994 by Vaidman \cite{Vaidman94}. The latter protocol utilizes the two-mode squeezed vacuum as the entangled resource; its major advantage is the principal capability of \emph{complete} transfer of a quantum state of an optical mode, independent, in particular, of the number of photons therein.

\begin{figure}[h]
\includegraphics[width=\columnwidth]{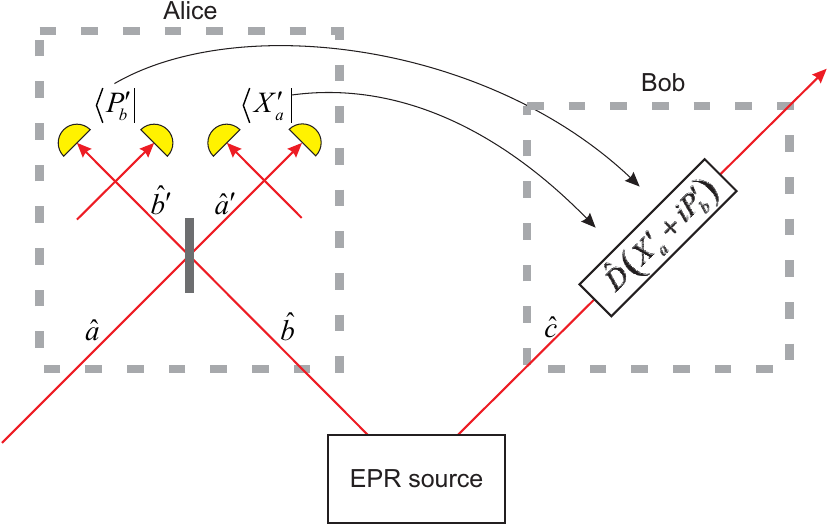}
\caption{\label{QTFig} The scheme of quantum teleportation. Operator $\hat D(X_m+iP_m)$ denotes phase-space displacement.}
\end{figure}

Figure \ref{QTFig} shows the scheme of the protocol. The sender, Alice, has the signal state she wishes to teleport in mode $\hat a$. In addition, she and the receiver, Bob, share a two-mode squeezed state in modes $\hat b$ and $\hat c$. In order to perform teleportation, Alice overlaps modes $\hat a$ and $\hat b$ on a symmetric beam splitter and preforms position and momentum measurements in its outputs using two homodyne detectors. She then communicates the results of her measurement to Bob via a classical channel. Bob performs phase-space displacement of mode $\hat c$ in accordance with that information, after which the state of this mode becomes identical to the initial state of mode $\hat a$.

In order to visualize the physics of teleportation, let us think of the signal state as a point $(X_a,P_a)$ in the phase space (neglect the uncertainty principle for a moment). Further, we assume the initial two-mode squeezing of modes $\hat b$ and $\hat c$ to be infinite: $X_b=X_c$ and $P_b=-P_c$, with both these quantities being completely uncertain. The beam splitter transformation, in accordance with Eqs.~\eqref{BSOpTrans} , makes the position in mode $\hat a$ equal to $X'_a=(X_a-X_b)/\sqrt 2$ while the momentum in mode $\hat b$ becomes $P'_b=(P_a+P_b)/\sqrt 2$.

Suppose now that the position and momentum measurements of these modes are performed, producing some specific results $X'_a$ and $P'_b$, respectively. This means that the position of mode $\hat b$ prior to the beam splitter has been $X_b=X_a-X'_a\sqrt 2$ and its momentum $P_b=-P_a+P'_b\sqrt 2$. Because of the infinite two-mode squeezing of modes $\hat b$ and $\hat c$ this implies, in turn, that $X_c=X_a-X'_a\sqrt 2$ and  $P_c=P_a-P'_b\sqrt 2$.

We see that, after Alice's measurement, the position and momentum of Bob's mode become certain and related to those of the initial state. Furthermore, if Alice communicates the observed values of $X'_a$ and $P'_b$ to Bob (via a classical channel), Bob will be able to perform a phase-space displacement operation (see Sec.~\ref{DispSec}) on his mode, obtaining the position and momentum equal to $X_a$ and  $P_a$, respectively, i.e. identical to those of the initial signal state.

A more rigorous argument can be presented in terms of Wigner functions. Let the initial Wigner function of the signal state be $W_a(X_a,P_a)$. The Wigner function of the EPR state shared between Alice and Bob is $W_{bc}(X_b,P_b,X_c,P_c)\propto\delta(X_b-X_c)\delta(P_b+P_c)$. The three-mode Wigner function is then $W_{abc}(X_a,P_a,X_b,P_b,X_c,P_c)=W_a(X_a,P_a)W_{bc}(X_b,P_b,X_c,P_c)$. After the beam splitter in Alice's channel, it will transform into
\begin{eqnarray}\label{Wabc}
W'_{abc}(X'_a,P'_a,X'_b,P'_b,X_c,P_c)&\propto& W_a\left(\frac{X'_a+X'_b}{\sqrt 2},\frac{P'_a+P'_b}{\sqrt 2}\right)\nna
&&\hspace{-4cm}\times\delta\left(\frac{-X'_a+X'_b}{\sqrt 2}-X_c\right)\delta\left(\frac{-P'_a+P'_b}{\sqrt 2}+P_c\right),
\end{eqnarray}
where the primed indices refer to the quadratures of the modes after the beam splitter. A measurement of $X'_a$ and $P'_b$ will yield, in mode $\hat c$,
\begin{eqnarray}\label{Wc}
W'_{c}(X_c,P_c)&=&\iint\limits_{-\infty}^\infty W'_{abc}(X'_a, P'_a,X'_b,P'_b,X_c,P_c)\de P'_a\de X'_b\nna
&\propto&W_a\left(X_c+X'_a\sqrt 2,P_c+P'_b\sqrt 2\right).
\end{eqnarray}
Again, applying displacement to Bob's mode, we recover a state with the Wigner function equal to that of the initial signal --- that is, the state identical to the initial.

In experimental practice, the teleportation performance is degraded by a number of factors, of which the primary ones are the optical losses, optical phase fluctuations and imperfect squeezing of the TMSV resource. A variety of performance metrics has been proposed \cite{Braunstein00,Ralph98,Grosshans02,Hammerer05,Braunstein05}. The most common one is the \emph{coherent-state fidelity}, which is the average, over all coherent states $\ket\alpha$, of the fidelity $F_c=\braketop\alpha{\hat T(\ketbra\alpha\alpha)}\alpha$, where $\hat T(\ketbra\alpha\alpha)$ is the density operator of the teleported state. For a perfect teleportation procedure, $F_c=1$. On the other hand, the best fidelity that can be achieved without the use of entangled resource, simply by Alice's measuring the position and momentum quadratures of the input state and Bob's recreating a coherent state with the same central position and momentum, is $F_c=1/2$.  The value of $F_c$ reaching a value of $2/3$, known as the no-cloning fidelity \cite{Grosshans02}, guarantees that nobody else can have a better copy of the input state than Bob. For this reason, the no-cloning fidelity is of relevance to continuous-variable quantum communication. The value of $2/3$ is also the minimum required for obtaining teleported states with negative values of the Wigner function.

The first continuous-variable quantum teleportation experiment was reported by Furusawa and colleagues in 1998 \cite{KimbleQT}. The TMSV resource has been obtained from two single-mode squeezed fields generated as counterpropagating modes in a single OPA cavity. Phase-space displacement was implemented using a low-transmissivity beamsplitter, with the amplitude and phase of the displacement beam regulated by electro-optical modulator. The resulting fidelity, $F_c=0.58$, exceeded the classical benchmark.

Thereafter, numerous efforts have been reported to refine the protocol and teleport increasingly complex quantum states. For example, Takei and colleagues \cite{Takei05} in 2005 demonstrated entanglement swapping (teleportation of one channel of a TMSV state), which is an essential component of quantum repeaters. Yonezawa \iea~\cite{Yonezawa07} teleported in 2007 a squeezed vacuum state and obtained, for the first time, squeezing in the output. The first teleportation of states with a negative Wigner function, such as the single photon and ``Schr\"odinger kitten" \cite{Ourjoumtsev06} was implemented in 2011 \cite{Lee11}. This work was followed by unconditional high-fidelity teleportation of dual-rail single-photon qubits \cite{Takeda13}.


\section{Applications in quantum metrology}
Squeezed light can be useful in any task that requires precise evaluation of the optical phase. Such tasks occur, for example, in optical communications \cite{Slavik2010} and metrology \cite{Schnabel2010}. Phase evaluation typically involves an interferometer, and its precision is determined by the phase uncertainty of the fields used. The coherent state, which is readily obtained from lasers, has a phase uncertainty on a scale of the inverse of its amplitude, or inverse square root of its photon number $1/\sqrt N$ [Fig.~\ref{SqWigFig}(b)].
However, employing nonclassical states has a potential to improve the precision up to the fundamental limit $\sim 1/N$ established by the Heisenberg uncertainty principle. Among the many approaches leading to this goal \cite{Giovannetti2011}, phase squeezing is perhaps the most straightforward [Fig.~\ref{SqWigFig}(f)]. In this section, we discuss a prominent example: the application of squeezed light in gravitational wave detection.

Gravitational waves (GWs) are  deformations of the space-time continuum caused by accelerating massive objects and propagating at the speed of light. GWs are a primary prediction of Einstein's general relativity, but they have not yet been observed due to their minuscule magnitude. The strongest GWs reaching the Earth are expected to cause deformations on a scale of 1 part in $10^{20}$, and their detection constitutes one of the most significant challenges faced by modern physics.

Gravitational wave detectors use a Michelson-type laser interferometer to detect small perturbations to positions of massive, freely suspended mirrors in its two arms. The action of a GW stretches one of the arms and compresses the other, thereby affecting the path-length difference and changing the intensity of the output signal.

\begin{figure}[h]
\includegraphics[width=\columnwidth]{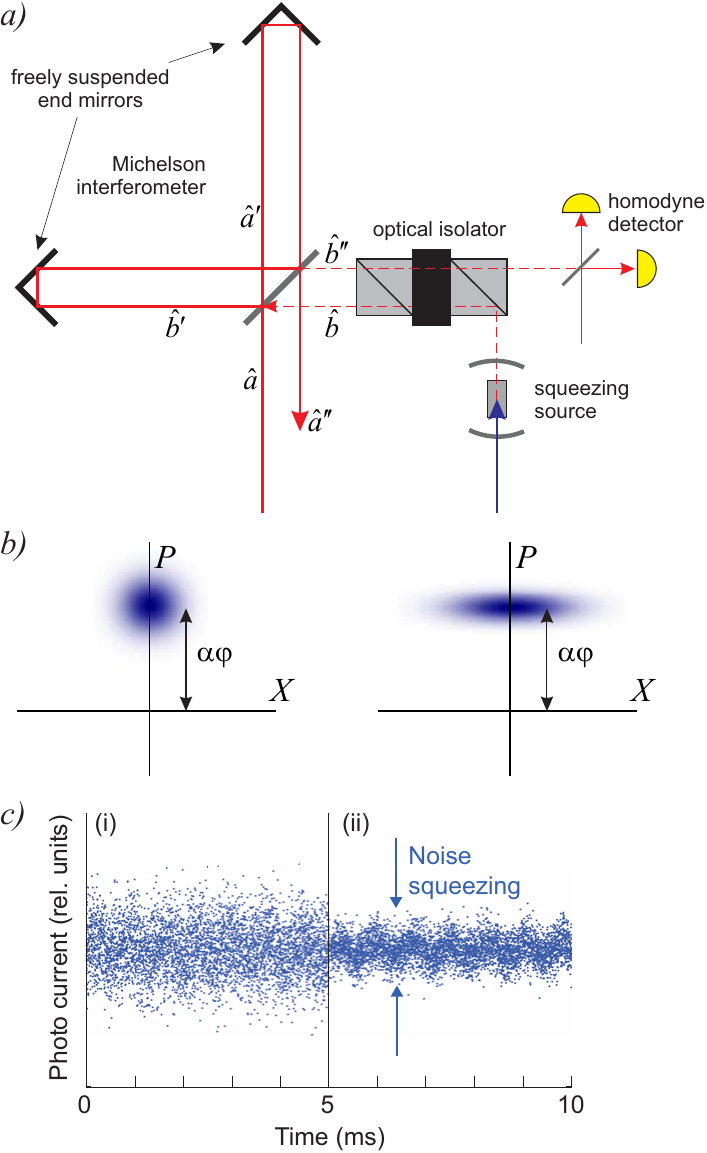}
\caption{\label{SqGMFig} Using squeezed vacuum to enhance the sensitivity of interferometric phase detection. a) Scheme of the setup. The squeezed vacuum is injected through the dark port (mode $\hat b$) of the interferometer. An optical isolator is used to separate the input and output modes $\hat b$ and $\hat b''$, and also to prevent scattering of the macroscopic light inside the interferometer into the squeezed vacuum source. For clarity, the end mirrors are sketched as retroreflectors. b) The Wigner function of the state in mode $\hat b''$ for the vacuum (left) and squeezed (right) input in mode $\hat b$. The momentum quadrature measurement by the homodyne detector is proportional to the interferometer path length difference; the measurement precision is enhanced by the initial momentum squeezing of mode $\hat b$. c) Simulation from Ref.~\cite{Schnabel2010} illustrating how squeezing helps revealing a small oscillation. Left, no squeezing; right, squeezing present.}
\end{figure}

At present, the world's most sensitive GW detectors are TAMA in Japan, GEO 600 in Germany, LIGO in the US and VIRGO in Italy. These detectors utilize a number of techniques in order to enhance their signal.
\begin{itemize}
\item The interferometer arms are constructed up to a few kilometers in length in order to increase the absolute displacement of the mirrors.
\item Both arms of the interferometer are turned into high-finesse Fabry-Perot cavities by means of additional mirrors placed near the Michelson interferometer beam splitter.
\item Massive (tens of kg) mirrors are used in order to reduce the radiation pressure noise and the mirrors' Brownian motion.
\item Laser powers of up to hundreds of watt are employed to minimize the phase uncertainty.
\end{itemize}
Further enhancement of any of these parameters would be prohibitive in terms of costs and resources. This is why additional sensitivity improvement associated with squeezing becomes useful. The idea of this improvement was proposed by Caves in 1981 \cite{Caves1981} and involves injecting squeezed vacuum into the dark port of the interferometer.

Suppose the interferometer  input mode $\hat a$ is fed with a strong laser field in coherent state $\ket\alpha$ [Fig.~\ref{SqGMFig}(a)]. We assume $\alpha$ to be real. The other input mode, $\hat b$, is in the vacuum or squeezed vacuum state. The beam splitter implements the mode transformation according to $\hat a'=(\hat a+\hat b)/\sqrt 2$, $\hat b'=(\hat a-\hat b)/\sqrt 2$. Let the interferometer paths be slightly unbalanced in length so that, upon return to the beam splitter, mode $\hat b'$ acquires a small phase shift $\varphi$ which we wish to evaluate. After interacting, for the second time, on the beam splitter, the modes become $\hat a'' =(\hat a'+\hat b'e^{i\varphi})/\sqrt 2$, $\hat b''=(\hat a'-\hat b'e^{i\varphi})/\sqrt 2$. Using $e^{i\varphi}\approx 1+i\varphi$, we find
\begin{equation}\label{IntGM}
\hat b''=\hat b-i\varphi\hat a.
\end{equation}
Because $\varphi$ is small, the second term in \eeqref{IntGM} effectively results in displacement of the (squeezed) vacuum mode $\hat b$ along the momentum axis by $\varphi\alpha$ [Fig.~\ref{SqGMFig}(b)].

A momentum quadrature measurement performed on mode $\hat b''$ by means of a homodyne detector will yield this value, with an uncertainly equal to the momentum uncertainty of the initial state of mode $b$. If this state is momentum squeezed, the measurement sensitivity is enhanced accordingly, as illustrated in Fig.~\ref{SqGMFig}(c).

The actual measurement procedure that is currently implemented in GEO 600 \cite{LIGO2011} and LIGO \cite{LIGO2013} largely follows the above description. A major challenge is to construct a source capable of generating squeezing in the frequency band compatible with gravitational waves. Typical GWs are produced in the audio range between 150 and 300 Hz, whereas most OPA-based squeezing sources built until recently exhibited significant technical noises at frequencies below 1 MHz. A series of breakthroughs achieved over the past decade helped identifying and eliminating the sources of these noises \cite{Schnabel2010}.

The primary issue turned out to be macroscopic optical field at the wavelength of the desired squeezing present within the OPA cavity. Mechanical fluctuations of the cavity length (which occur at low frequencies) randomly affect the magnitude and phase of that field and subsequently contaminate the output. The remedy consisted of preventing the ambient laser field from penetrating into the cavity. This included using a field of different frequency to lock the cavity length \cite{Vahlbruch2006}, using an optical isolator to prevent the reflection of the local oscillator from the homodyne detector photodiodes into the OPA cavity \cite{McKenzie2004} and even minimization of scattering from the nearby optical elements \cite{Vahlbruch2007}.

The most recent result on incorporating squeezed light into a GW detector has been reported for LIGO \cite{LIGO2013}. Enhancement of sensitivity of up to 2.2 dB for frequencies down to 150 Hz is reported. Note that this enhancement is far below the $>10$ dB degree of squeezing produced by the source employed. This is because of the losses introduced when injecting the squeezed field into the Michelson interferometer, imperfect mode matching with the carrier field, and phase fluctuations. It is expected that the next generation of LIGO (the so-called Advanced LIGO) will address most of these shortcomings \cite{LIGO2013}.

\section{Conclusion and outlook}
Over the past thirty years, the science of squeezed light has experienced enormous progress and made significant influence on the entire field of physics. Its primary effect, in my opinion, was to radically change the physicists' perception of quantum theory of electromagnetic radiation. Prior to the observation of squeezing, it was a largely abstract discipline, having little connection to experimental practice. Observation of squeezing and subsequent development of optical homodyne tomography resulted in techniques of creating, manipulating and measuring quantum states of light, allowing the postulates of quantum theory of light to be directly tested and applied in experiment.

The second important contribution of squeezing is that to  quantum information science. It provided an entangled resource for many quantum information protocols. Additionally, it gave rise to deeper understanding of parametric down-conversion, allowing preparation of other important quantum optical resources such as polarization-entangled photon pairs. As a result, optics has become, for at least a decade, the main test bed for quantum information science, effectively jump starting this field.

What developments can be expected in the next years? We are currently witnessing the emergence of new means of production of squeezing, e.g.~by bringing light into interaction with an optomechanical cavity, i.e. a optical cavity with one of its elements suspended so as to form on a high-quality mechanical resonator \cite{Safavi13,Purdy13}. The pressure of light inside the cavity on that resonator results in optical nonlinearities described by equations similar to \eqref{NLn}, thereby leading to the squeezing. The promise of this new method is the possibility to manufacture on-chip sources of squeezed light, enabling compact optical sensors and new fundamental tests of physics.

In terms of applications, major results are awaited in \emph{gravitational wave detection}. Although squeezed light has already been integrated into some of the detectors, it has not yet been used in actual data acquisition runs. In Advanced LIGO, the squeezing is expected to enhance the sensitivity by up to a factor of ten. Hopefully, such a detector will not only be able to prove the existence of GWs, but provide information about their spatial distribution and temporal dynamics. This would result in a fundamentally new method for observing the universe, which has a potential to revolutionize the entire field of astronomy.

No less exciting are squeezed light's contributions to quantum information science. Existing techniques of two-mode squeezing and quantum teleportation can be employed for the development of the \emph{continuous-variable quantum repeater} \cite{Briegel98}, which will dramatically enhance the quantum communication distance leading to global ``quantum internet". The unsolved challenges in this domain are long-term storage of squeezed light \cite{Honda08,Appel08,Lvovsky09} as well as methods of distilling the two-mode squeezed state that has experienced losses \cite{Takahashi10,Kurochkin13,Kurochkin13a}.

Recently, exciting developments have been reported on creation of \emph{multimode quadrature-entangled states} by simultaneous pumping of multiple spatial \cite{Armstrong12}, spectral \cite{Menicucci08,Pysher11}, or temporal \cite{Yokoyama13} modes of an SPDC arrangement. In this way, a large-scale, individually-addressable entangled state is created that may be possible to use in measurement-based quantum computation and other quantum information applications.



\bibliographystyle{osajnl}
\bibliography{SqReviewBib}

\end{document}